# Multimodal encoding in a simplified model of intracellular calcium signaling

Maurizio De Pittà[1], Vladislav Volman[2,3], Herbert Levine[2], Eshel Ben-Jacob[1,2,*]

1. School of Physics and Astronomy, Tel Aviv University, 69978 Ramat Aviv, Israel

2. Center for Theoretical Biological Physics, UCSD, La Jolla, CA 92093-0319, USA

3. Computational Neurobiology Lab, The Salk Institute, La Jolla, CA 92037, USA

*Corresponding author:

eshel@tamar.tau.ac.il,

Tel.: +972 3 640 7845

Fax: +972 3 642 5787

**Abstract**

Many cells use calcium signalling to carry information from the extracellular side of the plasma membrane to targets in their interior. Since virtually all cells employ a network of biochemical reactions for $Ca^{2+}$ signalling, much effort has been devoted to understand the functional role of $Ca^{2+}$ responses and to decipher how their complex dynamics is regulated by the biochemical network of $Ca^{2+}$-related signal transduction pathways. Experimental observations show that $Ca^{2+}$ signals in response to external stimuli encode information via frequency modulation or alternatively via amplitude modulation. Although minimal models can capture separately both types of dynamics, they fail to exhibit different and more advanced encoding modes. By arguments of bifurcation theory, we propose instead that under some biophysical conditions more complex modes of information encoding can also be manifested by minimal models. We consider the minimal model of Li and Rinzel and show that information encoding can occur by amplitude modulation (AM) of $Ca^{2+}$ oscillations, by frequency modulation (FM) or by both modes (AFM). Our work is motivated by calcium signalling in astrocytes, the predominant type of cortical glial cells that is nowadays recognized to play a crucial role in the regulation of neuronal activity and information processing of the brain. We explain that our results can be crucial for a better understanding of synaptic information transfer. Furthermore, our results might also be important for better insight on other examples of physiological processes regulated by $Ca^{2+}$ signalling.

**Keywords**: calcium; information encoding; astrocyte; bifurcation; Li-Rinzel

**Introduction**

The release of $Ca^{2+}$ ions from intracellular stores is a central event in the encoding of extracellular hormone and neurotransmitter signals (Berridge et al., 2000). In a multitude of different cells these signals impinge on G-protein coupled receptors on the cell membrane, which are linked to the inositol 1,4,5-trisphosphate ($IP_3$) intracellular pathway that triggers oscillations in the cytoplasmic $Ca^{2+}$ concentration (Hille, 2001). The level of stimulation determines the degree of activation of the receptor and therefore can be directly linked to the intracellular $IP_3$ concentration (Verkhratsky and Kettenmann, 1996). In turn, this latter process defines the type of intracellular $Ca^{2+}$ dynamics. One can therefore think of the $Ca^{2+}$ signal as being an encoding of information about the level of $IP_3$.

In recent years, a large amount of experimental observations showed that astrocytes, the main type of glia cells in the brain, can respond to synaptic activity by increases of their intracellular $Ca^{2+}$ levels (Dani et al., 1992; Porter and McCarthy, 1996; Parpura et al., 1994; Pasti et al., 1997; Wang et al., 2006). Evoked $Ca^{2+}$ oscillations in these cells can be confined locally within the same cell or, depending on the intensity of the stimulus, spread to other astrocytes (Sneyd et al., 1994, 1995; Charles, 1998). In addition, $Ca^{2+}$ oscillations in astrocytes induce release of several neurotransmitter types from these cells which feed back onto pre- and post- synaptic terminals, thus making astrocytes active partners in synaptic transmission (Araque et al., 1998).

Although the physiological meaning of $Ca^{2+}$ signaling in astrocytes remains currently unclear, a long-standing question has been how it could possibly participate in the encoding of synaptic information (Volterra and Meldolesi, 2005). Experimental evidence suggests that the frequency of astrocytic $Ca^{2+}$ oscillations is likely to be the preferential way of encoding of synaptic activity (Parpura, 2004). Increases in frequency or intensity of synaptic stimulation result in a corresponding increase in the frequency of $Ca^{2+}$ oscillations (Pasti et al., 1997). Notwithstanding, it is likely that encoding of synaptic activity in the astrocytic $Ca^{2+}$ signal could also occur in a more complex fashion (Carmignoto, 2000). Indeed, $Ca^{2+}$ oscillations can be highly variable in amplitude, depending on the intensity of the stimulation (Cornell-Bell et al., 1990; Finkbeiner, 1993), and their dynamics does not simply mirror the stimulation (Perea and Araque, 2005) (see also Figure 1).

From the modeling perspective, encoding of $Ca^{2+}$ oscillations is investigated by considering time-dependent solutions of systems of nonlinear differential equations as opportune biophysical parameters are changed time by time. It has been shown that amplitude modulations or frequency modulations of $Ca^{2+}$ oscillations can be reproduced separately by simple models consisting of two first-order differential equations (Dupont and Goldbeter, 1998; Li and Rinzel, 1994; Tang and Othmer, 1995). Notwithstanding, only those models which consider the diffusion of intracellular $Ca^{2+}$ or higher-order models with more equations have been acknowledged so far to exhibit different and more advanced encoding modes (Falcke, 2004).

In the present study we extend the investigations presented in (De Pittà et al., 2008) which demonstrated that the same cell could encode information about external stimuli either in amplitude modulation (AM) of calcium oscillations, or in frequency modulation (FM) or in both (AFM), by means of changes of its biophysical parameters. We utilize bifurcation theory to provide a general criterion for parameter tuning in minimal models that would allow the coexistence of amplitude and frequency encoding in $Ca^{2+}$ dynamics. Although discussed for the specific case of astrocytes, our results can be virtually extended to any cell type that displays some form of Ca$^{2+}$ signaling.

**Methods**

*A simplified description of Ca$^{2+}$ dynamics: the Li-Rinzel model*

We consider the Li-Rinzel (L-R) reduced version (Li and Rinzel, 1994) of the De Young-Keizer model for IP$_3$R kinetics (De Young and Keizer, 1992) as it represents a convenient compromise between generality and simplicity for the purposes of our study. The model assumes that periodic release of Ca$^{2+}$ ions from the endoplasmic reticulum (ER) can be brought about through the regulatory properties of the IP$_3$ receptor (IP$_3$R), the main type of ER calcium release channel in astrocytes and non-excitable cells in general (Lytton et al., 1992). Under these hypotheses, intracellular calcium balance is determined by only three fluxes, corresponding to: (1) a passive leak of Ca$^{2+}$ from the ER to the cytosol ($J_{leak}$); (2) an active uptake of Ca$^{2+}$ into ER ($J_{pump}$) due to the action of (sarco)-ER Ca$^{2+}$-ATPase (SERCA) pumps; and (3) a Ca$^{2+}$ release ($J_{chan}$) that is mutually gated by Ca$^{2+}$ and IP$_3$ concentrations (denoted hereafter as [Ca$^{2+}$] and [IP$_3$] respectively) (Figure 2).

The ER Ca$^{2+}$ pump rate is taken as an instantaneous function of [Ca$^{2+}$] and assumes the Hill rate expression with a Hill constant of 2:

$$J_{pump}\left([Ca^{2+}]\right) = \frac{v_{ER}[Ca^{2+}]^2}{K_{ER}^2 + [Ca^{2+}]^2}$$

where $v_{ER}$ is the maximal rate of Ca$^{2+}$ uptake by the pump and $K_{ER}$ is the SERCA Ca$^{2+}$ affinity, the concentration of Ca$^{2+}$ at which the pump will operate at half of its maximal capacity. The passive leakage current is assumed to be proportional to the

Ca$^{2+}$ gradient across the ER membrane by $r_L$, the maximal rate of Ca$^{2+}$ leakage from the ER:

$$J_{leak}([Ca^{2+}]) = r_L ([Ca^{2+}]_{ER} - [Ca^{2+}])$$

with [Ca$^{2+}$]$_{ER}$ being the Ca$^{2+}$ concentration inside the ER stores.

The release of Ca$^{2+}$ from the ER stores mediated by the IP$_3$R channel is also proportional to the Ca$^{2+}$ gradient across the ER membrane, but in this case the permeability is given by the IP$_3$R maximum permeability ($r_C$) times the channel's open probability. This latter is based on a gating model which assumes the existence of three binding sites on each IP$_3$R subunit: one for IP$_3$ and two for Ca$^{2+}$ which include an activation site and a separate site for inactivation. Therefore, three distinct gating processes are considered: IP$_3$ binding ($m_\infty$) and Ca$^{2+}$ binding to an activation site ($n_\infty$) and to a separate one for inactivation ($h$). Experimental data suggest a power of 3 for the opening probability (Bezprozvanny et al., 1991; De Young and Keizer, 1992), thus:

$$J_{chan}([Ca^{2+}], h, [IP_3]) = r_C m_\infty^3 n_\infty^3 h^3 ([Ca^{2+}]_{ER} - [Ca^{2+}])$$

where:

$$m_\infty = \frac{[IP_3]}{[IP_3] + d_1} \quad n_\infty = \frac{[Ca^{2+}]}{[Ca^{2+}] + d_5}$$

$J_{chan}$ and $J_{leak}$ can be grouped into $J_{rel}$ which represents the total Ca$^{2+}$ release flux from the ER:

$$J_{rel}([Ca^{2+}], h, [IP_3]) = J_{chan}([Ca^{2+}], h, [IP_3]) + J_{leak}([Ca^{2+}])$$
$$= (r_C m_\infty^3 n_\infty^3 h^3 + r_L)([Ca^{2+}]_{ER} - [Ca^{2+}])$$

Since the model considers the case of an isolated cell, namely Ca$^{2+}$ fluxes across the membrane are neglected, then the cell-averaged total free Ca$^{2+}$ concentration ($C_0$) is conserved and [Ca$^{2+}$]$_{ER}$ can be expressed in terms of equivalent cell parameters as $[Ca^{2+}]_{ER} = (C_0 - [Ca^{2+}])/c_1$ where $c_1$ is the ratio of the ER volume to the cytosolic volume. Thus $J_{rel}$ can be written entirely as a function of cell parameters as follows:

$$J_{rel}([Ca^{2+}], h, [IP_3]) = (r_C m_\infty^3 n_\infty^3 h^3 + r_L)(C_0 - (1+c_1)[Ca^{2+}])$$

Accordingly, the cytoplasmic $Ca^{2+}$ balance equation can be written as:

$$\frac{d}{dt}[Ca^{2+}] = J_{rel}([Ca^{2+}], h, [IP_3]) - J_{pump}([Ca^{2+}])$$

The above equation is combined with the second one for the kinetics of IP$_3$R/channel gating. By assuming fast IP$_3$ binding and $Ca^{2+}$ activation, the gating kinetics of the IP$_3$R can be lumped into a single equation for a dimensionless variable $h$ which represents the fraction of inactivated IP$_3$ receptors by cytoplasmic $Ca^{2+}$:

$$\frac{d}{dt}h = \frac{h_\infty - h}{\tau_h}$$

where:

$$h_\infty = \frac{Q_2}{Q_2 + [Ca^{2+}]} \quad \tau_h = \frac{1}{a_2(Q_2 + [Ca^{2+}])} \quad Q_2 = d_2 \frac{[IP_3] + d_1}{[IP_3] + d_3}$$

Values of parameters are provided in Table 1.

*Bifurcation analysis*

$Ca^{2+}$ dynamics is in equilibrium when both the cytoplasmic $Ca^{2+}$ level is constant, namely $\frac{d}{dt}[Ca^{2+}] = 0$, and the fraction of inactivated IP$_3$R does not change, i.e. $\frac{d}{dt}h = 0$.

In the phase plane $[Ca^{2+}]$ vs. $h$, the points for which $\frac{d}{dt}[Ca^{2+}] = 0$ trace the $Ca^{2+}$-*nullcline*. Analogously, $h = h_\infty([IP_3])$ constitutes the equation of the *h-nullcline*, as the points of such curves are solutions of $\frac{d}{dt}h = 0$. It follows that equilibrium points of the L-R equations coincide with the intersections of the nullclines (Figure 3a).

The shape of the two nullclines determines where and how they intersect, thus fixing the dynamical behavior of the model (Izhikevich, 2007; De Pittà et al., 2008). On the other hand, the nullcline shape depends itself on the choice of the model parameters, hence different parameter values can be associated with different kinds of $Ca^{2+}$ dynamics.

In the L-R model, the level of $IP_3$ is directly controlled by signals impinging on the cell from its external environment. In turn, one can get a complete picture of $Ca^{2+}$ dynamics at different levels of stimulation, by considering different $[IP_3]$ values and looking at the nullcline intersections.

For the original set of parameters, for $[IP_3]$ values in the physiological range of $0 < [IP_3] \leq 5$ μM (Parpura and Haydon, 2000), nullclines of the L-R equations always intersect in one point (Figure 3b). However, the stability of the equilibrium point depends on the $IP_3$ level, as shown in Figure 3a. At low $IP_3$ values ($[IP_3]=0.1$ μM) corresponding to basal conditions or weak stimulation, the equilibrium is stable, leading to constant $[Ca^{2+}]$. Such stability is then lost for higher $IP_3$ concentrations ($[IP_3]=0.5$ μM) when $Ca^{2+}$ oscillations rise in response to the external stimulus. Eventually, for higher $IP_3$ values ($[IP_3]=1.0$ μM), the equilibrium becomes stable again: in these conditions in fact, the system is in an over-stimulated resting state with the cytosolic $Ca^{2+}$ level that is kept high and constant by the strong stimulation.

We can summarize these observations by noting that the dynamics of the system dramatically changes, or *bifurcates,* when the equilibrium changes from stable to unstable and vice versa. Technically speaking, in such conditions the system is said to undergo a bifurcation, and since the $IP_3$ level triggers the bifurcation, it is referred to as the control (or bifurcation) parameter (Guckenheimer and Holmes, 1986).

We can map the bifurcations of the L-R system for the given set of parameters by plotting the equilibrium point, and analysing how it changes as $IP_3$ varies across its range of values (Figure 4a). In this way we find that at $[IP_3]=0.355$ μM, the system looses its stability via supercritical Hopf bifurcation. Namely, at this point $Ca^{2+}$ oscillations of arbitrarily small amplitude arise thanks to the appearance of a limit cycle in the phase plane (Figure 5a) (Rinzel and Ermentrout, 1989).

On the other hand, starting from $IP_3$ values as high as $[IP_3]=0.637$ μM oscillations are dampened to an overexcited steady $Ca^{2+}$ concentration through a subcritical Hopf bifurcation (Izhikevich, 2007), and dampening is faster as $[IP_3]$ increases (namely when the stimulus gets stronger).

Between the two bifurcations, the amplitude of the limit cycle increases as the level of $IP_3$ increases (Figure 4a) whereas the period (frequency) does not change significantly (Figure 4b). Hence, for the original set of parameter values, the L-R model encodes the information about the level of $IP_3$ by amplitude modulations (AM) of $Ca^{2+}$ oscillations (see also Figure 15a).

*Linking bifurcations with physiology*

The emergence of $Ca^{2+}$ oscillations in the L-R model can be interpreted in terms of physiology once we consider the steady calcium fluxes for different $IP_3$ values as determined by setting $h$ to $h_\infty(C)$. These fluxes capture the fast time scale response of the system close to the fixed point, since the rate of inactivation of receptors is slower than that of $Ca^{2+}$ release. We note that stable $Ca^{2+}$ levels are found either for basal or very low $IP_3$ concentrations and for high values of this latter, when the fixed point is at high $Ca^{2+}$ concentrations, and the slope of the efflux $J_{rel}$ is negative (Figure 5b). In these conditions $J_{chan}$ is vanishing so that intracellular $Ca^{2+}$ levels are fixed by the balance between $J_{leak}$ and $J_{pump}$. On the contrary, instability and Ca$^{2+}$ oscillations, occur for IP$_3$ values such that there exist values of [Ca$^{2+}$] where $J_{rel} < J_{pump}$ and the slope of $J_{rel}$ is positive, namely $J_{chan}$ is increasing. In these conditions, the cytosolic Ca$^{2+}$ concentration is also increasing until the slope of $J_{rel}$ becomes negative and a return flow in the phase plane towards smaller Ca$^{2+}$ values is observed. However, at the same time the slope of $J_{rel}$ returns positive and the process repeats so that self-sustained Ca$^{2+}$ oscillations occur.

The dynamical flow does not exhibit a strong positive amplification as it is essentially due to the slow inactivating dynamics of the IP$_3$ receptor channels. This accounts for the above observed independence of the frequency from the IP$_3$ level, as the frequency of the oscillations is close to that of the *h* time delay, which indeed does not vary much across the oscillatory regime.

**Results**

*FM dynamics*

In light of several other experimental findings that reported frequency modulations (FM) of Ca$^{2+}$ oscillations in response to variations of the level of IP$_3$ (Pasti et al., 1997), we re-examined the L-R model to investigate if and how changes in the biophysical parameters could lead to frequency-modulated dynamics (De Pittà et al., 2008). We found that either a higher Ca$^{2+}$ affinity of SERCA pumps, $K_{ER}$, or a lower IP$_3$ affinity of the receptor channels, $d_5$, or a smaller rate of Ca$^{2+}$ leakage from the ER, $r_L$, all can regulate the switching between AM and FM encoding dynamics.

Under such conditions, the Ca$^{2+}$-nullcline becomes sharply N-shaped (Figures 6a,b) so that a narrow interval of IP$_3$ values appears for which nullclines intersect in three points, whereas outside such interval, their intersection consists of one point only. At [IP$_3$]=0.479 μM, a saddle and an unstable points appear in addition to the pre-existing stable node, through a saddle-node bifurcation (SN). As the concentration of IP$_3$ is increased, the saddle moves towards smaller Ca$^{2+}$ values and larger $h$ values, thus getting closer to the stable node (Figure 7). At [IP$_3$]=0.526 μM the stable point and the saddle coalesce via a saddle-node on invariant circle (SNIC) bifurcation and the only remaining attractor is a limit cycle (Izhikevich, 2007). Although the limit cycle emerges via a fold bifurcation related to a subcritical Hopf bifurcation of the stable node that precedes the SNIC (Figure 6c) bifurcation, the same limit cycle becomes invariant at the SNIC so that Ca$^{2+}$ oscillations rise at arbitrarily small frequency which is very sensitive to the distance from the bifurcation point (Figure 6d). On the contrary, the amplitude of the limit cycle remains almost constant and is considerably larger with respect to the AM dynamics (Figure 8a).

This fact can be understood by considering the calcium fluxes for a sample FM dynamics, as shown in Figure 8b. We note that a change in one of the above mentioned parameters, such as a higher $d_5$ value, shifts the intersection between $J_{rel}$ and $J_{pump}$ to lower Ca$^{2+}$ values. It follows that the range of IP$_3$ values for which there is instability is widened as the intersection between $J_{rel}$ and $J_{pump}$ occurs in coincidence with a positive slope of $J_{rel}$. Accordingly, the CICR is maximized and the rising phase is shortened. A faster increase of Ca$^{2+}$ is also lagged by a stronger SERCA uptake flux, so that the falling time of the Ca$^{2+}$ oscillation decreases too, and their shape becomes alike to spikes.

*Bifurcation approach to the characterization of the transition between AM and FM dynamics*

Although AM and FM encoding $Ca^{2+}$ dynamics have been observed distinctly, it is highly plausible that they coexist (Carmignoto, 2000). As we explain later, the major information-related effect of such coexistence is that it would enable simultaneous transfer of different kinds of information from the extracellular space to diverse intracellular targets.

From a mathematical point of view, the coexistence of AM and FM encoding in a minimal model such as the L-R model may be unexpected, given the incompatible nature of the bifurcations underlying these two modes. Notwithstanding, if we consider parameters such as $d_5$, $r_L$ or $K_{ER}$ that can make $Ca^{2+}$ dynamics to switch from AM to FM and vice versa, we then can inspect the transition between these two modes. In the parameter space we can predict that as the transition occurs, the dynamics keeps the essential feature of AM encoding (a great variability in the amplitude of oscillations as [$IP_3$] varies), but it is sufficiently close to FM encoding to also inherit from this latter the great variability in the frequency of oscillations (Figures 9,10).

By comparison of the bifurcation diagram of AM encoding (Figure 4a) with that of FM encoding (Figure 6c) we notice two substantial differences. First, two additional bifurcations of saddle-node type are found in the FM case with respect to AM. Second, the supercritical Hopf bifurcation in the AM encoding system which occurs at low [$IP_3$] levels, becomes subcritical in the FM case. Since the two saddle-node-like bifurcations cannot appear independently, given the associated nullcline shape when these two bifurcations exist, i.e. the existence of one/three intersections, it is clear that FM encoding occurs only when these two bifurcations appear together. Provided that saddle-node bifurcations are codimension-1 bifurcations, since they are triggered by changes of a single control parameter ([$IP_3$] in our case), it follows that their appearance must be through a codimension-2 bifurcation. Namely, a further control parameter, different from [IP$_3$], must be considered in order for these bifurcations to appear. Indeed, in order to obtain FM encoding and the associated two saddle-node bifurcations, we need to consider the bifurcation diagram with IP$_3$ as a control parameter and see how such diagram changes as a second control parameter like $d_5$, $r_L$ or $K_{ER}$ changes. In our case the codimension-2 bifurcation that "adds" two saddle-node-like bifurcations to the system is a Cusp bifurcation (Kuznetsov, 1998) (Figure 11a). Accordingly, any parameter like $d_5$, $r_L$, or $K_{ER}$ which can trigger the L-R-model through a Cusp bifurcation, can switch Ca$^{2+}$ dynamics from AM to FM encoding modes and back.

Furthermore, when in the FM mode, the saddle appears through the saddle-node bifurcation at lower IP$_3$ values, two heteroclinic trajectories are born (Edelstein-Keshet, 1988). Such trajectories, which constitute the unstable manifold of the saddle, leave the saddle to end into the preexisting stable node. As previously stated, in FM encoding, oscillatory dynamics emerges through a subcritical Hopf bifurcation at IP$_3$ values that are close but lower than the IP$_3$ concentration at the SNIC. We can state that such Hopf bifurcation must be subcritical and must precede the Cusp (Figure 12). We can check the validity of these argumentations by inverse reasoning: if the Hopf bifurcation was supercritical as in the AM case, a limit cycle of infinitesimal amplitude would appear at the bifurcation point. Then, for higher IP$_3$ values before the SNIC, such a limit cycle would increase in amplitude, crossing with the two heteroclinic trajectories of the unstable manifold. Accordingly, in the phase plane there would exist two points both belonging to two different trajectories: one constituted by the limit cycle and the other by the heteroclinic orbits. This would contradict the fundamental existence-uniqueness theorem (Perko, 2001). Therefore the Hopf bifurcation at lower IP$_3$ values must be subcritical when FM encoding dynamics occurs.

Similar arguments to those used to explain the nature of the Cusp bifurcation, imply that a supercritical Hopf bifurcation can turn into subcritical, through a codimension-2 Bautin bifurcation (Kuznetsov, 1998) (Figure 11b). Moreover, given that the appearance of FM dynamics requires the existence of a subcritical Hopf bifurcation at low IP$_3$ values, it follows that from AM to FM, the supercritical Hopf bifurcation must change into subcritical before the two saddle-node bifurcations appear. In other words, the system must undergo a Bautin bifurcation before the Cusp bifurcation occurs, that is equivalent to saying that the existence of a Bautin bifurcation on the Hopf point at low [IP$_3$] hints the existence of the Cusp bifurcation. From another point of view, if a parameter in an AM-encoding version of the L-R model can drive the supercritical Hopf point (H+) through a Bautin bifurcation, then it is possible that the same parameter makes the system capable of FM encoding through a Cusp bifurcation. That is, the existence of a Bautin bifurcation for the Hopf point at low [IP$_3$] in AM encoding is a *necessary* condition for the possible occurrence of a Cusp bifurcation and the transition of the dynamics towards FM encoding.

If we reverse the direction of the transition, namely from FM to AM, such a condition becomes also *sufficient*. Indeed, when for any of the free parameters (except for [IP$_3$] and the one used to obtain FM encoding) the Hopf point at low [IP$_3$] levels - now subcritical - undergoes a Bautin bifurcation, no saddle-node bifurcation can take place otherwise as above demonstrated, a violation of the existence-uniqueness theorem would occur. It follows that a Cusp bifurcation, which accounts in this case for the *disappearance* of the two saddle-node bifurcations, must occur before the Bautin bifurcation. This is equivalent to saying that in a transition from FM to AM encoding, the Cusp bifurcation anticipates the Bautin bifurcation, namely the occurrence of the Cusp bifurcation is a necessary condition for the existence of the Bautin bifurcation. In other words, if a Bautin bifurcation occurs while changing a parameter in the FM encoding mode, then it is as the system was closer to AM encoding.

*Coexistence of amplitude and frequency modulations: the AFM dynamics*

A different reasoning accounts for the amplitude of the oscillations. When a Hopf bifurcation point undergoes a Bautin bifurcation, a fold limit cycle bifurcation also occurs simultaneously (Izhikevich, 2007). Then, going from AM to FM, the preexisting limit cycle that was generated through a supercritical Hopf bifurcation, now splits into two limit cycles, the external one being stable, and the inner one unstable. As [$IP_3$] level increases, the inner unstable limit cycle shrinks to the unstable focus originated through the Hopf bifurcation, whereas the external limit cycle tends to coalesce with the two closing heteroclinic trajectories of the unstable manifold of the saddle, eventually leading to the homoclinic orbit at the basis of the SNIC. Because the subcritical Hopf point in FM-encoding conditions is very close to the SNIC point, the amplitude of the FM oscillations is essentially fixed by the amplitude of the homoclinic orbit. In general the amplitude of such homoclinic orbit is at least 2-fold wider than the orbits that rise through supercritical Hopf bifurcation in the AM case. Therefore, when going from AM to FM, the gap must be filled starting from the Bautin bifurcation where, due to the simultaneous occurrence of a Hopf bifurcation with a fold limit cycle bifurcation, the amplitude of the external stable limit cycle becomes independent of the bifurcation parameter and can easily converge to the homoclinic orbit. Accordingly, the farther the Bautin bifurcation is from the appearance of the homoclinic orbit, the greater is the variability of the amplitude of oscillations. In other words, the farther the Bautin bifurcation is from the Cusp bifurcation, the better AM dynamics is displayed by the model.

In conclusion, if parameters of an AM-encoding version of the L-R system are tuned such that the Hopf point at low IP$_3$ concentration is in proximity of a Bautin bifurcation, then it is likely that Ca$^{2+}$ dynamics could show frequency encoding, as it "feels" the ghost of a close Cusp bifurcation. Conversely, in order to keep AM encoding (i.e., increasing amplitude with increasing stimulation), such Bautin bifurcation must be "far enough" from the Cusp bifurcation so that the limit cycle that originates from it and tends to the homoclinic orbit at the SNIC, could sweep within a range of amplitudes sufficiently large to guarantee AM encoding. If both these conditions can be simultaneously satisfied, that is if we can feel the ghost of the Cusp but we are sufficiently far from it with the Bautin that has already occurred, then the emerging oscillations would increase both in amplitude and frequency as [IP$_3$] increases (De Pittà et al., 2008). We will hereafter refer to this additional encoding mode where AM and FM coexist as "AFM encoding". Noteworthy is that AFM dynamics can be generally associated with a bifurcation diagram of the L-R model displaying two subcritical Hopf bifurcations only, whereas the opposite does not generally hold.

*A simple criterion for the coexistence of AM and FM dynamics: the CPB scenario*
In order to quantify how "sufficiently far" the Bautin bifurcation should be from the Cusp bifurcation in order to obtain AFM dynamics, we can estimate their reciprocal distance in terms of units of the control parameter that is responsible for their occurrence, and normalize it with respect to the value of the control parameter at the first bifurcation that is encountered in the transition from one dynamics to the other.

Accordingly, when going from AM to FM, the Bautin bifurcation precedes the Cusp bifurcation, so that such normalized distance can be written as:

$$\delta \text{BCP} = \left| \frac{\text{B} - \text{CP}}{\text{B}} \right|$$

Conversely, in the case of the opposite transition, from FM to AM, the same distance can be thought of as:

$$\delta \text{CPB} = \left| \frac{\text{CP} - \text{B}}{\text{CP}} \right|$$

If we assume that FM dynamics occurs when the period of oscillations doubles (halves) as the bifurcation parameter is changed, whereas AM encoding is consistent with an amplitude range (for stable oscillations) greater than the maximal amplitude of the oscillations that could possibly be displayed in the AM-encoding case, then AFM encoding is likely to be found if and only if $0.2 < \delta \text{CPB} < 0.8$ (or $0.2 < \delta \text{BCP} < 0.8$).

All these considerations can be summarized in what we can describe as Cusp-Bautin (CPB) scenario according to which within a Li-Rinzel-like dynamical system, a necessary condition for the coexistence of amplitude and frequency encoding, is that the tuning of a second parameter in addition to the $IP_3$ concentration changes the bifurcation diagram so that only two Hopf bifurcations exist and the one at low $[IP_3]$ is subcritical. This condition is also sufficient if it is possible to estimate the CPB (BCP) factor such that it is: *$0.2 < \delta \text{CPB} < 0.8$ ($0.2 < \delta \text{BCP} < 0.8$)*.

This scenario represents a heuristic criterion for the choice of parameters to tune the dynamics of a set of nonlinear equations of the L-R type. Hence it can help to reveal the existence of AFM encoding in other $Ca^{2+}$ pathways or even other intracellular signaling transduction pathways. From this perspective, the CPB scenario can also provide meaningful insights on the inner molecular machinery that could endow cells with a rich $Ca^{2+}$ dynamics, characterized by AM-, FM- or AFM-encoding oscillations.

*Cellular parameters as triggers of multiple encoding*

Results for our set of free parameters are summarized in Table 2. With respect to the original set of parameter values, the conditions of the CPB scenario, can only be fulfilled by a reduction of $IP_3R$ affinity for $Ca^{2+}$ activation such that $0.107<d_5<0.161$ µM. Nonetheless, AFM $Ca^{2+}$ dynamics can be found in numerous other combinations of parameters either in presence of low $IP_3R$ affinity for $Ca^{2+}$ activation ($d_5 = 0.2$ µM), or small $Ca^{2+}$ leakage rate from the ER ($r_L = 0.002$ s$^{-1}$) or high $Ca^{2+}$ affinity of SERCA pumps ($K_{ER} = 0.051$ µM). In all these cases in fact, a higher total cell-averaged free $Ca^{2+}$ concentration (Figures 11a,b;13), as well as a reduced maximal SERCA pump rate can provide AFM encoding. The same conditions are also found when $d_5$ is high or $K_{ER}$ is small and the leakage rate is higher with respect to the original value of $r_L = 0.11$ s$^{-1}$. Furthermore, in the case of high $d_5$ values only, AFM dynamics also occurs with $0.085<K_{ER}<0.127$ µM. In contrast, for low $K_{ER}$ values AFM encoding is found with a reduced $IP_3R$ affinity for $Ca^{2+}$ activation such as $0.085<d_5<0.127$ µM. Eventually, parameters like $c_1$ and $r_C$, namely the ratio between ER volume and cytosol volume and the maximal CICR rate respectively, seem not to affect the encoding features of $Ca^{2+}$ dynamics.

**Discussion**

Experimental evidence over the past decades indicates that $Ca^{2+}$ signals in response to external stimuli can encode information via frequency modulation (FM) as well as through amplitude modulations (AM) (Figure 15) (Berridge, 1997). Although both types of dynamics have been observed separately, it is likely that they coexist (Carmignoto, 2000). Nonetheless the physical bases for such coexistence are not yet understood.

In the present work we considered the Li-Rinzel minimal model for intracellular $Ca^{2+}$ dynamics and by arguments of bifurcation theory we showed that AM and FM encoding can coexist for opportune choices of parameter values (Figure 17).

From a mathematical perspective, such choices of parameters can be characterized by focusing on the backbone of codimension-2 bifurcations that are responsible for the dynamical transition of the model from AM to FM and vice versa.

On the other hand, from the perspective of physiology, our study suggests a critical role in $Ca^{2+}$ dynamics for the total free $Ca^{2+}$ level of the cell (Figure 14), as well as for the rate of $Ca^{2+}$ uptake by SERCA pumps in presence either of enhanced SERCA $Ca^{2+}$ affinity, or weak receptor $IP_3$ affinity or slow $Ca^{2+}$ leakage from the ER stores. Accordingly, these results provide strong theoretical support to the notion that more than the existence of complex feedback loops not included in the Li-Rinzel equations (Mishra and Bhalla, 2002), the primary source of the richness of intracellular $Ca^{2+}$ responses is constituted by heterogeneities of inherent cell's properties (Toescu, 1995).

Our discussion considered only ideal deterministic calcium dynamics. In real cells stochastic fluctuations can play an important role and strongly affect the dynamical behavior. In particular, it has been suggested that many cell types can exhibit spontaneous oscillations even below the Hopf bifurcation, due to random openings of clusters of small numbers of $IP_3Rs$ (Callamaras and Parker, 2000; Shuai and Jung, 2003). Indeed stochastic calcium events occurring in the absence of external stimulation were observed in astrocytes (Nett et al., 2002). Since the rate of such openings depends on the IP$_3$ concentration, the cell could exhibit a noisy form of FM encoding even without saddle-node structure. Noteworthy is that such a possibility does not contradict our results; rather, it further supports the idea that the type of encoding carried out by Ca$^{2+}$ signals can be regulated through activation of intracellular mechanisms that control specific physiological parameters.

In the context of communication theory, AM and FM encoding of information has been extensively studied. More recently, mixed A/F modulation has received renewed interest as it has advantages in various tasks such as the coordination of informational input from multiple channels (Ono et al., 1999; Luo et al., 2006). Of course, these systems operate by modulating a carrier signal which is always active, even when nothing is being transmitted. In contrast, intracellular Ca$^{2+}$ signals uses a principle which can be regarded as *bifurcation-based* encoding. Here, we mean that the baseline IP$_3$ level is set to be sufficiently close to a bifurcation point so that opportune variations of IP$_3$ due to external signals regularly cross this point switching on or off the encoding of the stimulus (Figure 16). In the AM dynamics, the peak value of calcium encodes the information on the level of IP$_3$ which is directly linked to the strength of the stimulus impinging on the cell. In the FM case, variations in the IP$_3$ will trigger bursts of Ca$^{2+}$ spikes with information encoded in inter-spike intervals. In the mixed AFM mode (Figure 17), both features contain information which can be separately decoded by different downstream effectors with different Ca$^{2+}$ responses. This is particularly suitable for those systems that require special constraints in coordination of informational input from multiple channels, for which mixed AM and FM modulation (MM) has recently been receiving much attention (Sasha et al., 2001).

In particular, the above can be very relevant for the understanding of the role of astrocytes in the regulation of synaptic transmission. Astrocytes respond to synaptic activity through their intracellular $Ca^{2+}$ dynamics, which in turn feeds back to neurons by triggering the release of gliotransmitter (Bezzi et al., 2004; Parpura, 2004; Wang et al., 2006). Motivated by this scenario, Volman et al. (Volman et al., 2004) and Hulata et al. (Hulata et al., 2004, 2005) proposed that $Ca^{2+}$ dependent feedbacks from astrocytes onto synapses could be responsible of the long-range time ordering of recorded neuronal activity (Segev et al., 2002; Segev et al., 2004). Accordingly, neuronal network activity would result from a cooperative regulation by both neurons and glia so that networks should be viewed as neuron-glia fabrics rather than stand-alone neuronal webs.

Numerous other theoretical investigations further support this possibility. It was suggested that strong coupling between neurons and astrocytes over-expressing glutamate receptors could give rise to seizure-like spontaneous neural oscillations similar to those observed in epileptic tissue (Nadkarni and Jung, 2003). The unique statistical properties of inter-spike intervals of the firing activity of highly-active neurons could reliably be reproduced by modulation of synaptic activity by astrocytic AM-encoding $Ca^{2+}$ dynamics (Volman et al., 2007a). Furthermore $Ca^{2+}$-dependent astrocyte feedback might influence the stochasticity of synaptic transmission (Volman et al., 2007b; Nadkarni et al., 2008) thus linking astrocyte $Ca^{2+}$ signaling with plasticity and learning (Nadkarni and Jung, 2007).

We note that it is reasonable to expect that the inclusion of AFM encoding could have deep consequences. In these conditions, short-time scale effectors, mostly sensitive to the number of pulses and therefore associated with FM dynamics, are likely to be involved in feedback to the local synapses. On the contrary, long-time scale effectors, which integrate the total signal and hence are linked to AM dynamics, are likely to coordinate information with other astrocytes via intercellular signaling. The coexistence of multimodal encoding by astrocytic $Ca^{2+}$ signals is therefore candidate to influence the timing of the neuronal activity in a complex fashion, hinting the existence of astrocyte-regulated higher level information processing based on hidden principles in information coding and decoding, yet to be deciphered.

**Acknowledgements**

The authors thank V. Parpura, G. Carmignoto, B. Ermentrout, B. Sautois and N. Raichman for insightful conversations on $Ca^{2+}$ dynamics and its capability of encoding information. V. Volman acknowledges the support of U.S. National Science Foundation I2CAM International Materials Institute Award, grant DMR-0645461. This research has been supported by the Tauber Fund at Tel Aviv University, by the Maguy-Glass Chair in Physics of Complex Systems, and by he NSF-sponsored Center for Theoretical Biological Physics (grant nos. PHY-0216576 and PHY-0225630).

**Tables**:

| | Parameter | Value | |
|---|---|---:|---|
| $C_0$ | Total free $Ca^{2+}$ concentration (referred to cytosol volume) | 2 | µM |
| $c_1$ | ER/cytoplasm volume ration | 0.185 | -- |
| $r_C$ | Maximal CICR rate | 6 | $s^{-1}$ |
| $r_L$ | $Ca^{2+}$ leakage rate | 0.11 | $s^{-1}$ |
| $v_{ER}$ | Maximum SERCA uptake rate | 0.9 | $µMs^{-1}$ |
| $K_{ER}$ | SERCA pump activation constant | 0.1 | µM |
| $d_1$ | $IP_3R$ dissociation constant ($IP_3$) | 0.13 | µM |
| $d_2$ | $IP_3R$ dissociation constant ($Ca^{2+}$ inactivation) | 1.049 | µM |
| $d_3$ | $IP_3R$ dissociation constant ($IP_3$) | 0.9434 | µM |
| $d_5$ | $IP_3R$ dissociation constant ($Ca^{2+}$ activation) | 0.08234 | µM |
| $a_2$ | $IP_3R$ binding rate for $Ca^{2+}$ inactivation | 0.2 | $µMs^{-1}$ |

**Table 1**: Original set of parameter values for the Li-Rinzel model.

| Parameter | | $d_5$=0.2 µM | | | $r_L$=0.002 s$^{-1}$ | | | $K_{ER}$=0.051 µM | | |
|---|---|---|---|---|---|---|---|---|---|---|
| | | δCPB | AFM-Range | | δCPB | AFM-Range | | δCPB | AFM-Range | |
| $d_5$ | µM | 0.296$^*$ | 0.107 | 0.161 | 0.130 | | | 0.236 | 0.085 | 0.127 |
| $C_0$ | µM | 0.401 | 2.949 | 4.424 | 0.737 | 2.115 | 3.172 | 0.418 | 3.037 | 4.555 |
| $c_1$ | -- | -- | -- | -- | -- | -- | -- | -- | -- | -- |
| $r_L$ | s$^{-1}$ | 0.355 | 0.175 | 0.262 | 0.837$^*$ | -- | -- | 0.384 | 0.178 | 0.266 |
| $r_C$ | s$^{-1}$ | ∞ | -- | -- | 93.760 | -- | -- | ∞ | -- | -- |
| $v_{ER}$ | µMs$^{-1}$ | 0.250 | 0.149 | 0.595 | 0.332 | 0.139 | 0.555 | 0.274 | 0.144 | 0.577 |
| $K_{ER}$ | µM | 0.202 | 0.134 | 0.201 | 0.129 | -- | -- | 0.182$^*$ | -- | -- |

**Table 2**: AFM-encoding ranges according to the CPB scenario for three different cases of cellular properties. Cases distinguished by ($^*$) are based on the computation of δBCP rather than of δCPB, as the transition which they are associated to is from AM to FM. All the other cases are evaluated instead starting from FM conditions (according to parameter values specified in column labels).

**Figure Captions**:

**Figure 1**. (*a*) Purified astrocytes from rat's visual cortex loaded with calcium indicator fluo-3 and stimulated by norepinephrine (25 mM) and (*b*) associated normalized $Ca^{2+}$ traces. Similarly to astrocytes, many other cells display intracellular $Ca^{2+}$ oscillations in response to several external agents. Such oscillations are characterized by great variability both in their peaks and in their frequency, hinting complex modes of encoding of information of external signals. Images and experimental data courtesy of V. Parpura. Images and experimental data courtesy of V. Parpura [adapted from: Lee W, Parpura V (2007) Exocytotic release of glutamate from astrocytes: comparison to neurons. In: Bean A (ed) Protein trafficking in neurons. Elsevier, Amsterdam, pp 329–365]

**Figure 2**. Schematics of the Li-Rinzel model. The model considers a single cell in a $Ca^{2+}$-free extracellular environment, so that $Ca^{2+}$ fluxes across the membrane are neglected. Accordingly, intracellular $Ca^{2+}$ dynamics is elicited by $IP_3$ which is initially required to open the $IP_3$ receptors located on the ER membrane and to sensitize the channels toward feedback activation by cytoplasmic calcium. Following, calcium dynamics is controlled by the interplay of calcium-induced calcium release (CICR), a nonlinear amplification process regulated by the calcium-dependent opening of channels to ER calcium stores, and by the action of the active SERCA pumps, which enable a reverse flux. Basal $Ca^{2+}$ levels result instead from the balance of a nonspecific passive $Ca^{2+}$ leakage from the ER stores into the cytoplasm and active uptake by SERCA pumps.

**Figure 3** (colors online). Phase plane of the Li-Rinzel model. *(b)* For the original set of parameter values the $Ca^{2+}$-nullcline (*orange*) intersects the h-nullcline (*green*) always in one point *(a)* so that a bifurcation occurs only when the equilibrium from stable (•) becomes unstable (○) or vice versa.

**Figure 4**. Bifurcation diagram of the Li-Rinzel model. *(a)* For the original set of parameter values self-sustained $Ca^{2+}$ oscillations are born via supercritical Hopf bifurcation at $[IP_3]=0.355$ μM and die via subcritical Hopf bifurcation at $[IP_3]=0.637$ μM. Between these two values, the amplitude of the oscillations increases as $[IP_3]$ increases, whereas *(b)* their period is almost constant. Hence information about the level of $IP_3$ is encoded in the amplitude of the oscillations but not in their frequency, so that the model shows AM-encoding. Stable (unstable) oscillations are depicted by "•" or "○" respectively, at their maximum and minimum values.

**Figure 5** (colors online). *(a)* In the phase plane, oscillatory $Ca^{2+}$ dynamics can be associated with the existence of a (stable) limit cycle. *(b)* From a physiology perspective instead, oscillations rises at the onset of CICR when $J_{rel}$ (*black curve*) increases due to the increase of $J_{chan}$ (see text). In such conditions, the balance at basal conditions between $J_{pump}$ (*magenta curve*) and $J_{leak}$ is temporarily lost and the action of SERCAs lags the total release of $Ca^{2+}$ from the ER stores, till the cytoplasmic $Ca^{2+}$ level is high enough and inactivation of $IP_3$-gated channels takes over. Then $J_{pump}$ prevails on $J_{rel}$ and the original conditions are restored.

**Figure 6**. *(a, b)* An $IP_3$ affinity of the receptor as low as $d_5=0.2$ μM, makes the $Ca^{2+}$-nullcline sharply N-shaped so that this latter intersects with the h-nullcline in one or three points. *(c)* The change in the number of equilibrium points is associated with two saddle-node (SN) bifurcations, and the one at high [IP$_3$] level occurs on an invariant circle (SNIC). *(d)* For this reason oscillations emerge with arbitrarily small frequency but almost constant amplitude. Hence the model shows FM-encoding of the stimulus, as the information of the level of IP$_3$ is encoded in the frequency of the oscillations rather than in their amplitude.

**Figure 7** (colors online). Pictorial scheme of the rise of oscillations in FM dynamics. *(a)* At low [IP$_3$] level a saddle-node bifurcation occurs out of which a saddle ($\triangle$) and an unstable point ($\circ$) are generated. In these conditions, the unstable manifold of the saddle is the one drawn in magenta whereas the stable manifold is shown in blue and moves along the middle branch of the Ca$^{2+}$ nullcline. As [IP$_3$] level increases, a limit cycle appears via a fold limit cycle bifurcation and soon after the only stable point ($\bullet$) looses its stability through a subcritical Hopf bifurcation. *(b)* Accordingly, the unstable manifold of the saddle wraps around the limit cycle but two trajectories are still recognizable, namely a short branch that originates from the saddle towards decreasing Ca$^{2+}$ values and a longer one that leaves the saddle, makes and ample excursion in the region of high Ca$^{2+}$ values and eventually tends to the limit cycle. *(c)* As [IP$_3$] level further increases, the short branch of the stable manifold shrinks as the saddle and the unstable (focus) points coalesce. Then the two branches of the unstable manifold form a closed (homoclinic) orbit, which is responsible for the rise of oscillations with infinite period.

**Figure 8**. FM oscillations. *(a)* With respect to Figure 5a, in FM conditions the excursion drawn in the phase plane by the limit cycle is wider and its shape almost independent of the level of IP$_3$. *(b)* In terms of Ca$^{2+}$ fluxes, the shape of relaxation-

like oscillations in the FM dynamics can be understood once we note that higher $d_5$ values make $J_{rel}$ to intersect $J_{pump}$ with a positive slope at lower $Ca^{2+}$ concentrations. In such conditions the CICR is maximized and the rising phase is shortened as the range of $IP_3$ values for which $J_{rel}$ increases is wider. Since higher $Ca^{2+}$ concentrations are reached, a stronger SERCA uptake also takes place so that the falling time of the $Ca^{2+}$ oscillations decreases too and the overall oscillations resemble $Ca^{2+}$ spikes.

**Figure 9**. $Ca^{2+}$ encoding as a function of cellular parameters. Variations in the $IP_3$ affinity of the receptors ($d_5$) and in the level of cell-averaged total free $Ca^{2+}$ concentration ($C_0$), can account for different $Ca^{2+}$ dynamics: from AM encoding to FM encoding and their coexistence too. This is possible since the effect of a lower $d_5$, namely the shift of the dynamics towards FM, is counteracted by and increase in $C_0$. *(a)* $C_0$=2.0 µM *(b)* $C_0$=3.5 µM *(c)* $C_0$=5.0 µM. Columns, from left to right: $d_5$=0.08324 µM (original L-R value), $d_5$=0.15 µM, $d_5$=0.2 µM.

**Figure 10**. Period of oscillations according to bifurcations diagrams in figure 9.

**Figure 11**. Beyond the transition between AM and FM. *(a, b)* With respect to AM dynamics for the original set of parameter values (dashed curve), the bifurcation diagram of the FM-encoding Li-Rinzel model shows two additional saddle-node bifurcations (SN, SNIC) and a subcritical Hopf bifurcation (H-) at low [$IP_3$] level. *(a)* The SN and SNIC bifurcations appear together through a Cusp bifurcation at $d_5^{CUSP}=0.173$ µM ([$IP_3$]=1.157 µM). *(b)* The supercritical Hopf bifurcation (H+) in the original Li-Rinzel model instead becomes subcritical via Bautin bifurcation at $d_5^{BAUTIN}=0.134$ µM ([$IP_3$]=0.551 µM). As explained in the text, the prerequisite for the coexistence of AM and FM dynamics is a CPB factor between 0.2 and 0.8. Here $\delta\text{BCP} = (d_5^{BAUTIN} - d_5^{CUSP})/d_5^{BAUTIN} = 0.296$ which accounts for AFM dynamics for values of $d_5$ between 0.107 and 0.161 µM (see also Table 2). Wider AFM ranges can also be obtained by considering other parameters. *(c)* By fixing $d_5$=0.2 µM, and changing the cell-averaged total free Ca$^{2+}$ concentration, $C_0$, for example, we find that the two saddle nodes bifurcations disappear through a Cusp at $C_0^{CUSP} = 2.458$ µM ([IP$_3$]=0.944 µM). *(d)* Accordingly, the Hopf bifurcation at low [IP$_3$] level undergoes a Bautin bifurcation at $C_0^{BAUTIN} = 3.442$ µM ([IP$_3$]=0.396 µM). In these conditions $\delta\text{CPB} = (C_0^{CUSP} - C_0^{BAUTIN})/C_0^{CUSP} = 0.401$ which is consistent with a $C_0$ range within 2.949 and 4.424 µM for the occurrence of AFM dynamics.

**Figure 12** (colors online). The Hopf bifurcation at low [IP$_3$] must be subcritical in FM conditions. *(a)* A pictorial representation of the phase plane for 1.769<[IP$_3$]<2.172 µM at $d_5$=0.2 µM, when nullclines intersect in three points which are a stable point (•), a saddle (∆) and an unstable point (°). In these conditions the unstable manifold of the saddle (*magenta*) terminates into the stable point, whereas the stable manifold (*blue*) is along the middle branch of the Ca$^{2+}$-nullcline. *(b)* At [IP$_3$]=2.172 µM the stable point looses its stability through a subcritical Hopf bifurcation. If such bifurcation was supercritical then a limit cycle would appear around the equilibrium point and it would expand as [IP$_3$] increases. Then the limit cycle would cross with the unstable and stable manifold (◊) of the saddle thus violating the fundamental existence-uniqueness theorem and leading to an absurdity.

**Figure 13** (colors online). The space of coexistence. *(a)* A tridimensional rendering of the surface of equilibrium points as a function of $d_5$ (for $C_0$=2.0 µM) with superimposed curves of saddle-node bifurcations (*blue*) and Hopf bifurcations (*green*) (see also figures 11a,b). *(b)* The same surface as a function of the cell-averaged total free $Ca^{2+}$ concentration $C_0$.

**Figure 14** (colors online). Coexistence of amplitude and frequency modulations. Changes in dynamical properties may occur as a result of modulation of cell-average total free calcium concentration. *(a)* Dependence of oscillation amplitude on the value of $IP_3$, for different concentrations of the overall calcium for $d_5$=0.2 µM and *(b)* associated oscillation period in the same conditions. Red (FM): $C_0$=2 µM; black (AFM): $C_0$=4 µM; blue (AM): $C_0$=5 µM.

**Figure 15**. AM and FM encoding. The information on the intracellular level of $IP_3$ (as depicted in *(c)*) can be encoded through oscillations in the intracellular $Ca^{2+}$ level of the cell. *(a)* For the original set of parameter values such oscillations occurs at almost constant frequency whereas their amplitude changes according to the level of $IP_3$. Hence, in such a case, the information about the level of $IP_3$ is encoded in "Amplitude modulations" (AM) of $Ca^{2+}$ oscillations. *(b)* For $d_5$=0.2 µM instead, oscillations occurs at almost constant amplitude but their frequency depends now on $IP_3$ concentration. Accordingly, the information about the $IP_3$ stimulus is encoded in "Frequency modulations" (FM) of $Ca^{2+}$ oscillations.

**Figure 16**. Bifurcation-based encoding. The baseline $IP_3$ level is close to a bifurcation which is responsible for the rise of oscillations. Hence variations of [$IP_3$] make the system to cross such bifurcation switching encoding on or off. Notwithstanding if the stimulus level varies too fast with respect to the time scales of $Ca^{2+}$ release from the ER stores and receptor inactivation, no effective encoding can be observed (*left*) and both AM and FM dynamics are reduced to subthreshold-like perturbations of the resting cytosolic $Ca^{2+}$ level. This situation changes instead when the $IP_3$ is beyond the bifurcation point long enough to trigger oscillations (*right*). AM and FM parameters according to figure 15.

**Figure 17**. AFM encoding. As discussed in the text and summarized in Table 2, there might be several conditions related to inherent properties of the cell under which both AM and FM dynamics occurs simultaneously. Such AFM dynamics encodes the information of the $IP_3$ level (in *(d)*) in $Ca^{2+}$ oscillations whose amplitudes are associated each with a specific frequency (as shown in *(c)*).

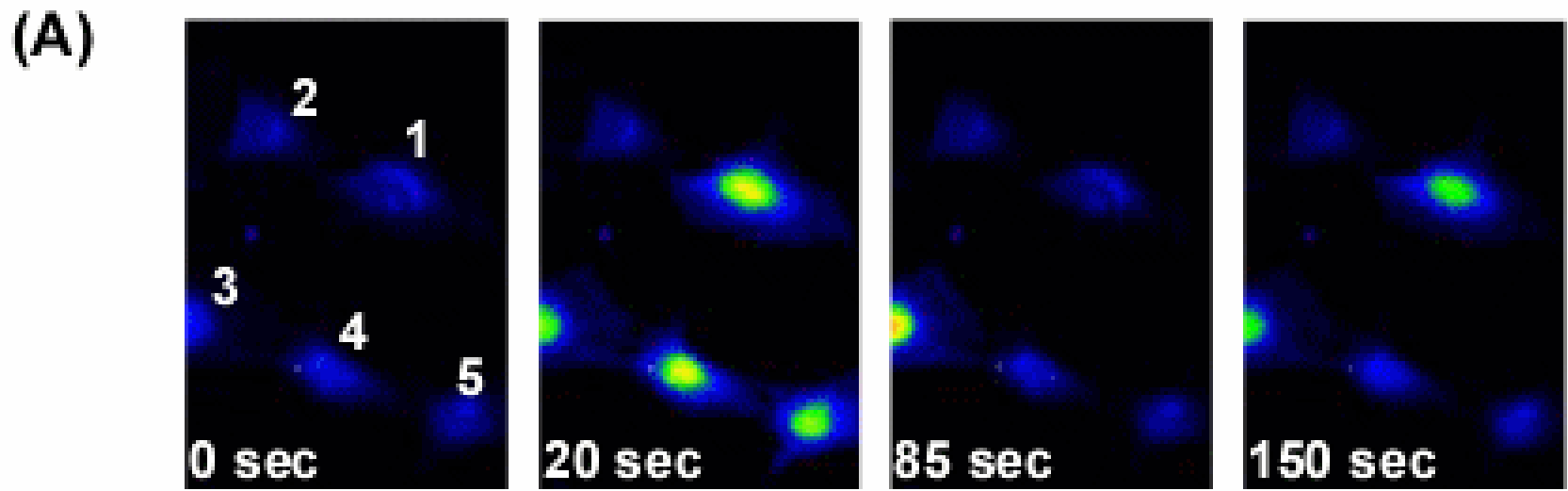
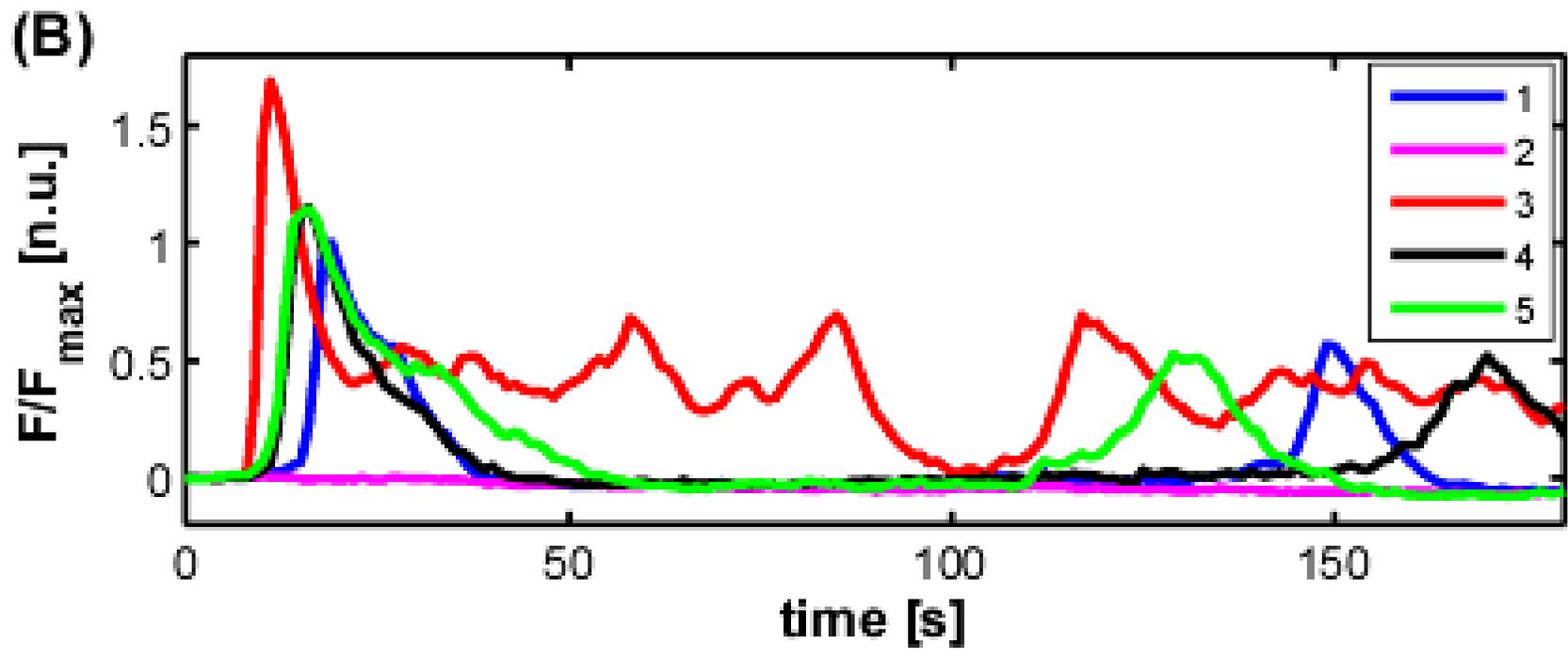

Figure 1

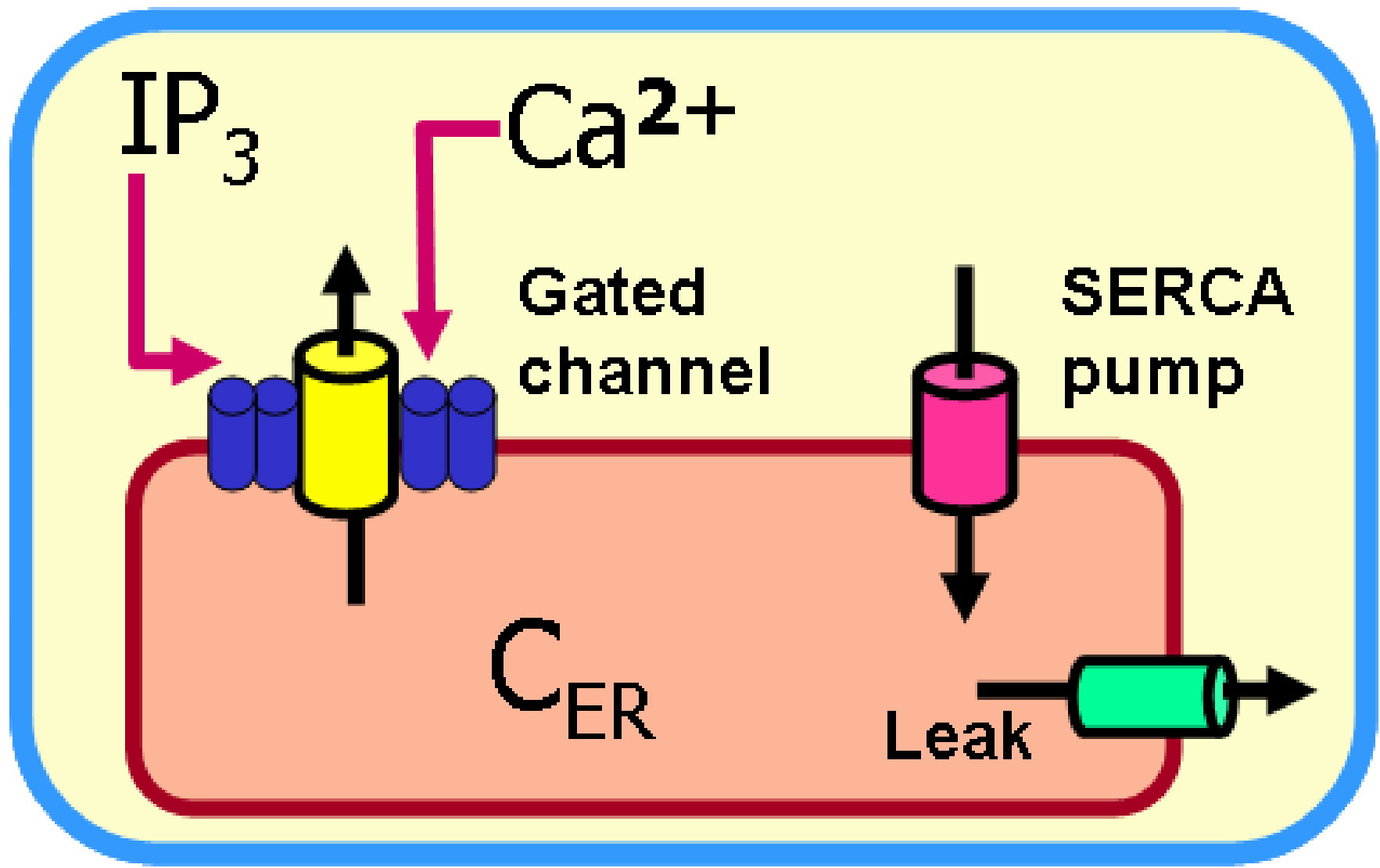

Figure 2

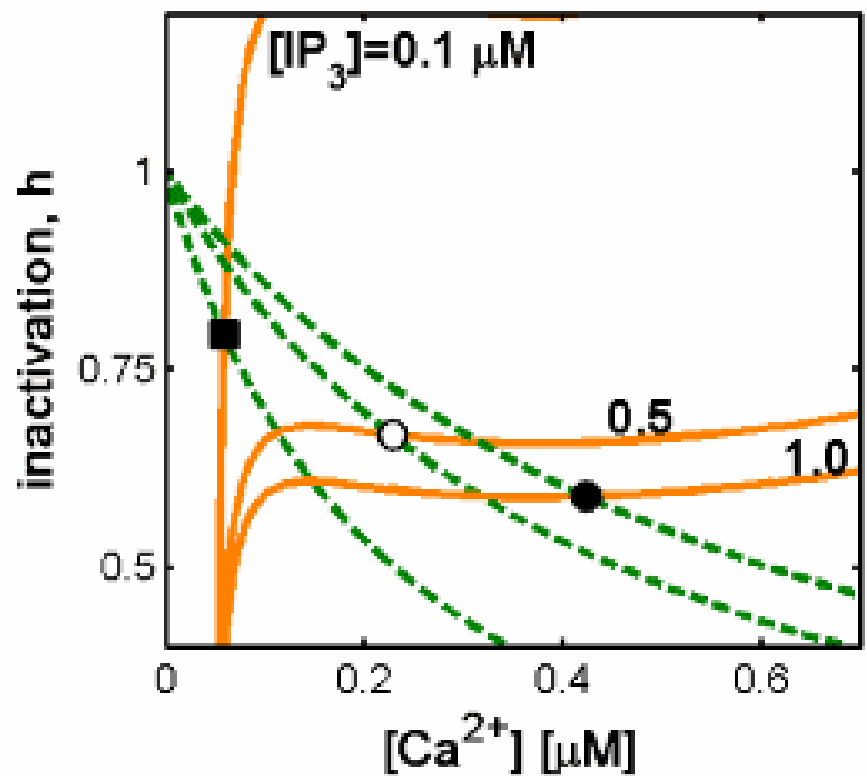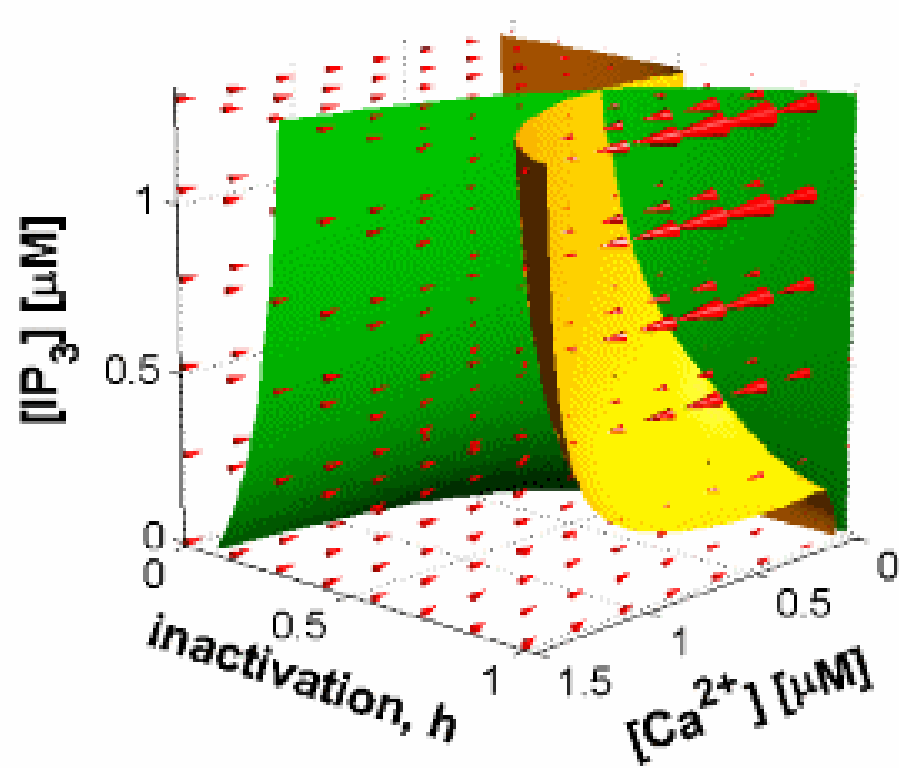

Figure 3

(A) 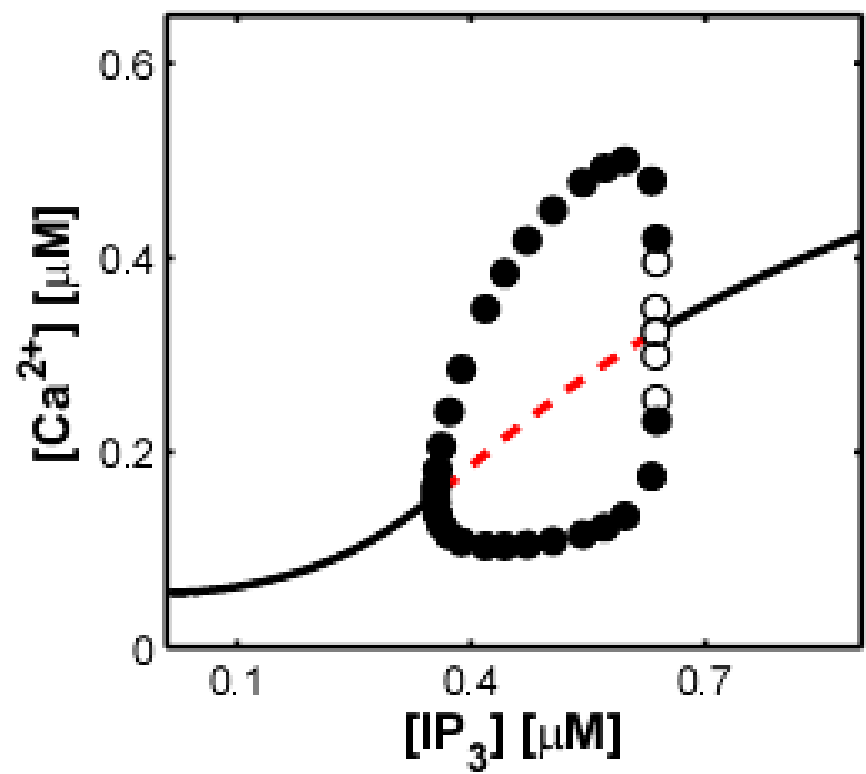
(B) 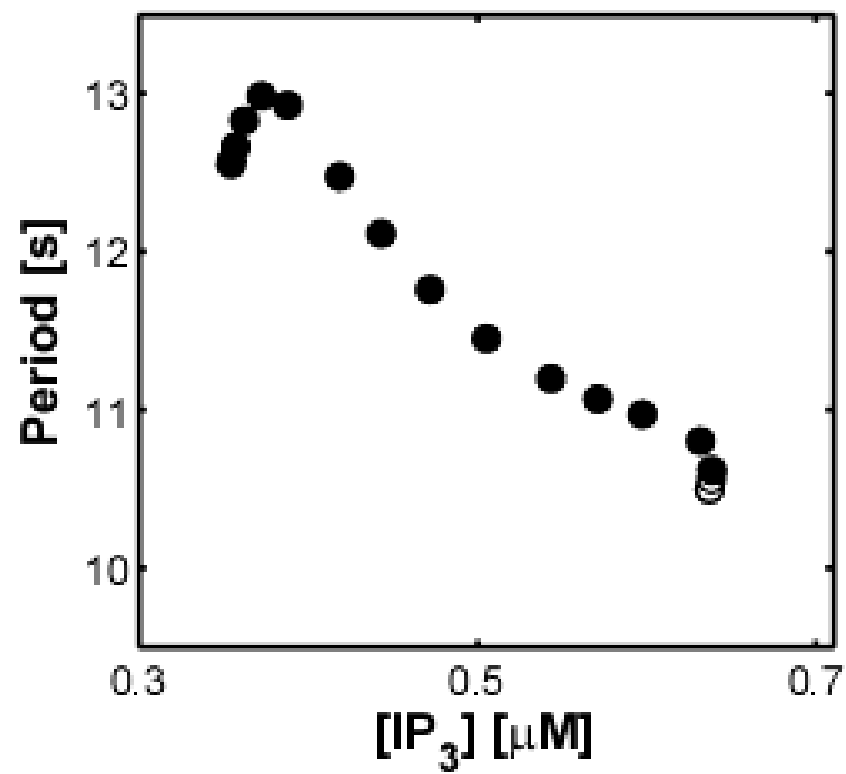

Figure 4

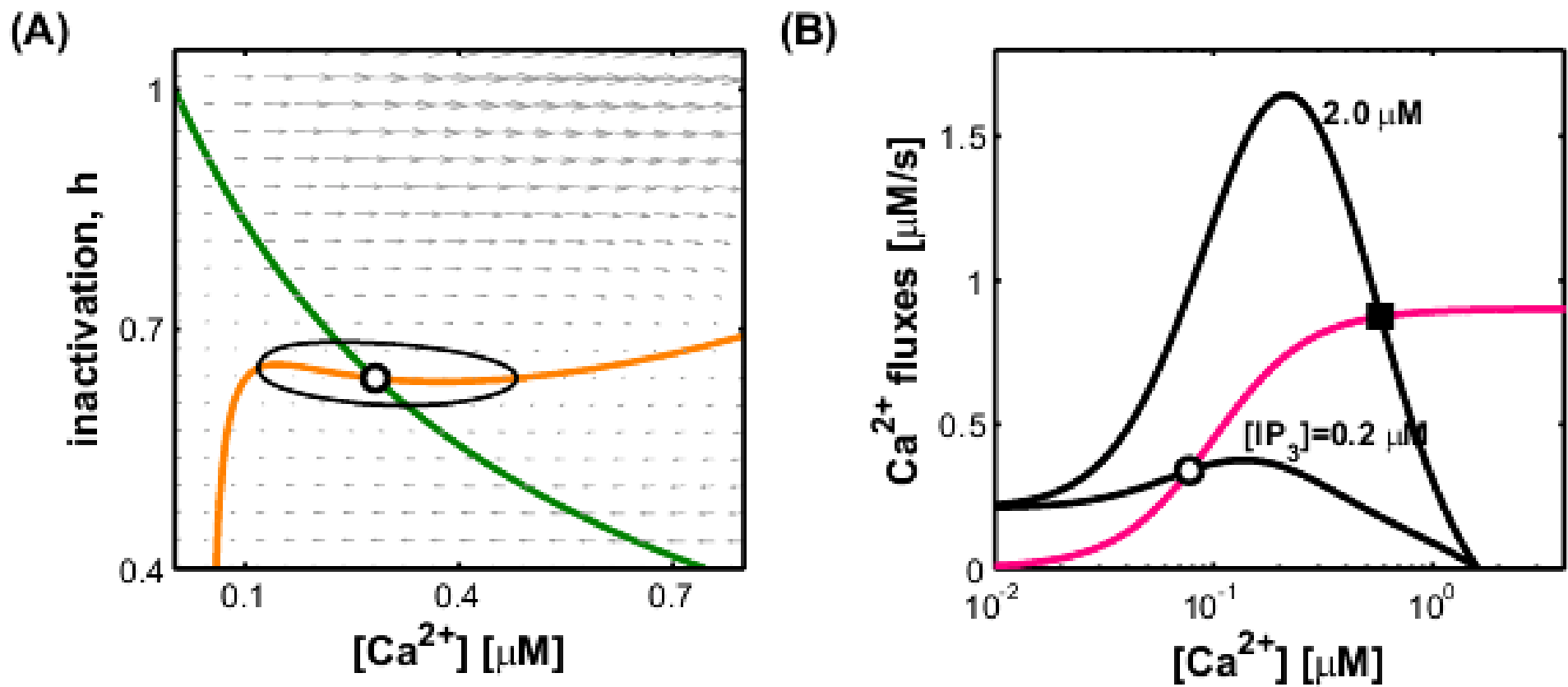

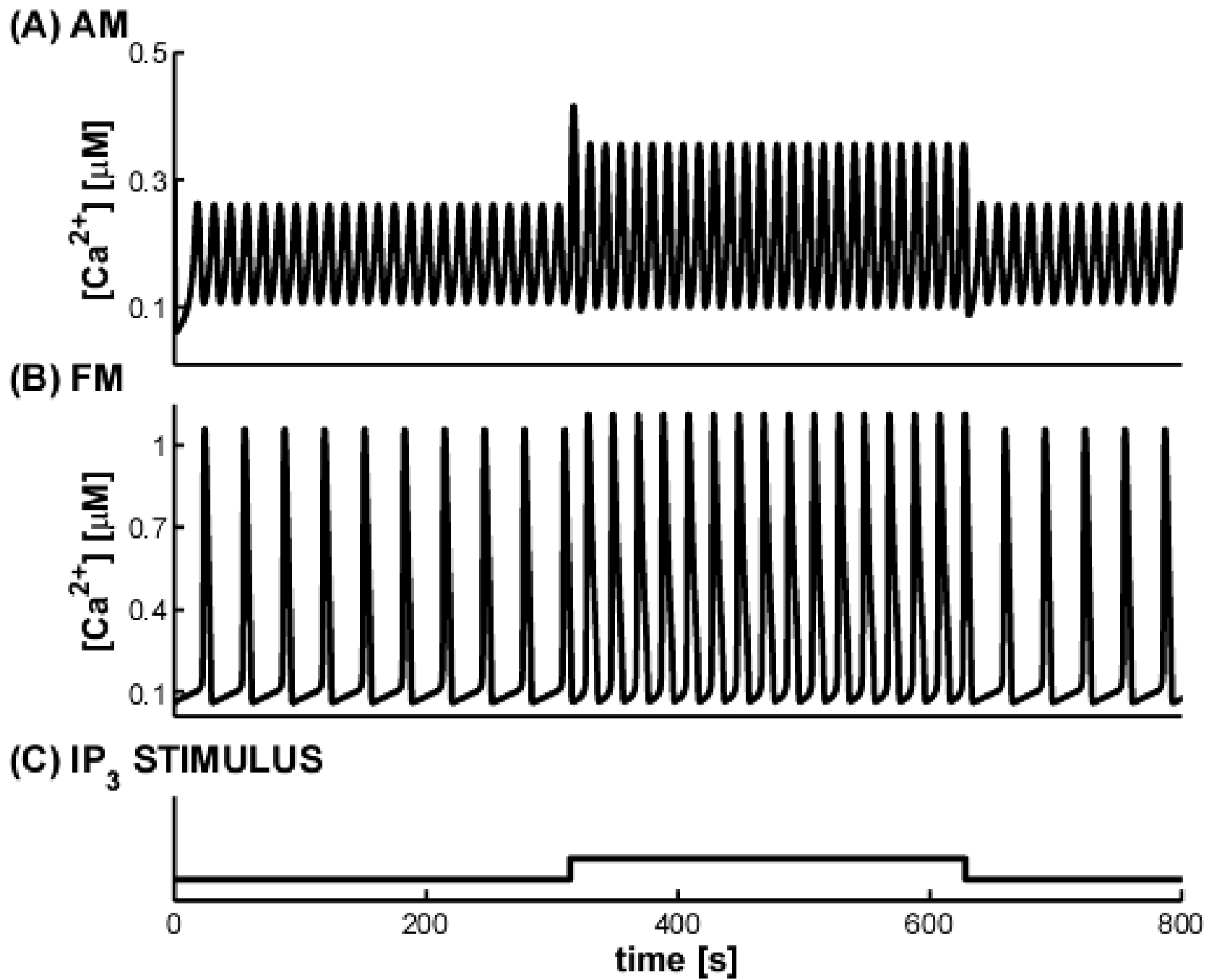

Figure 6

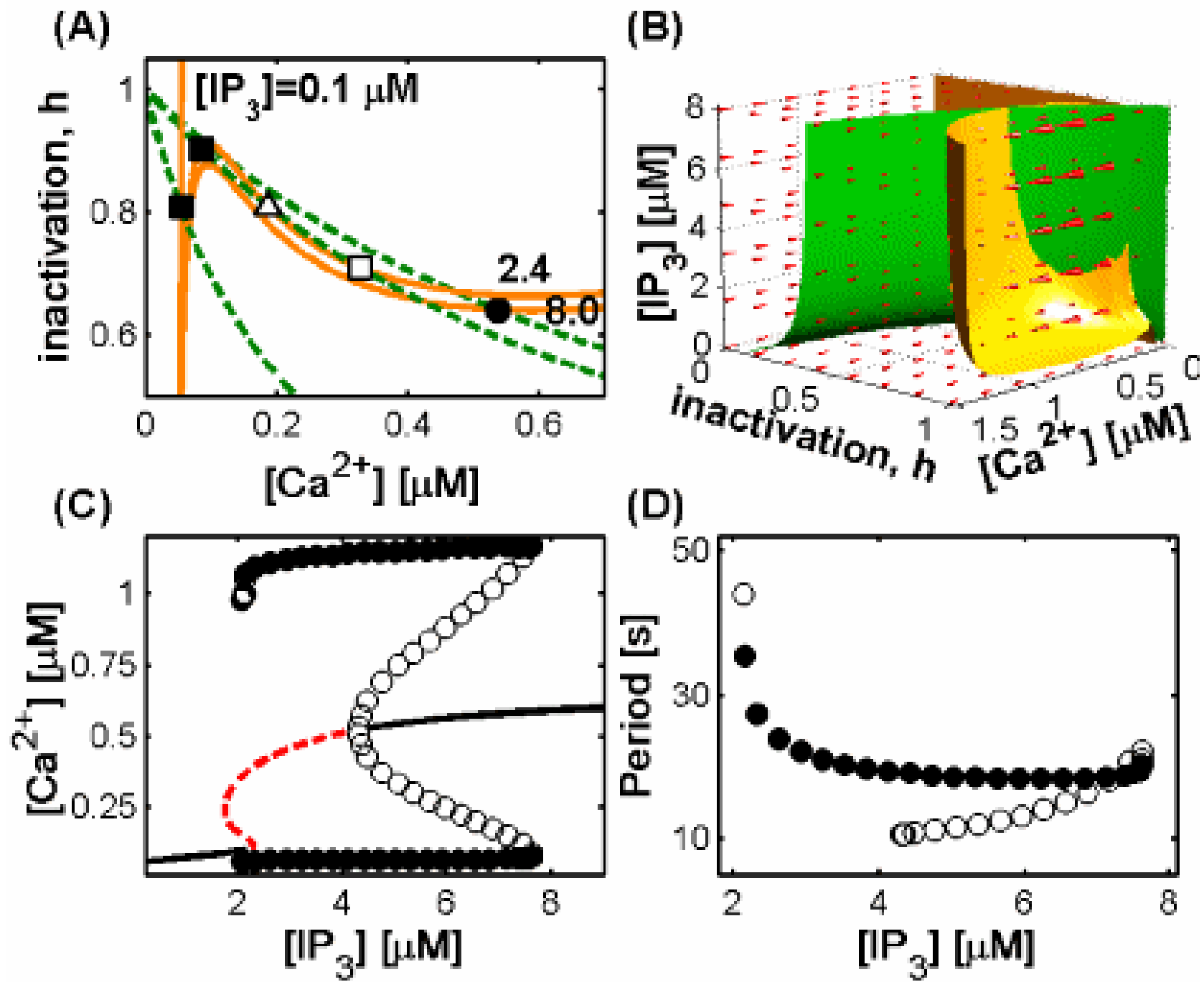

Figure 7

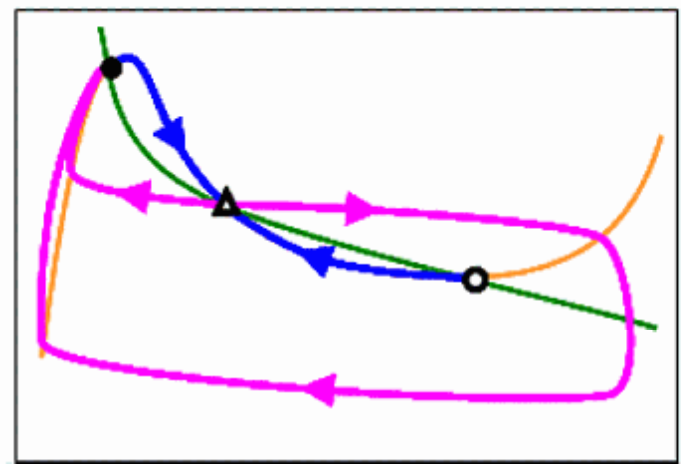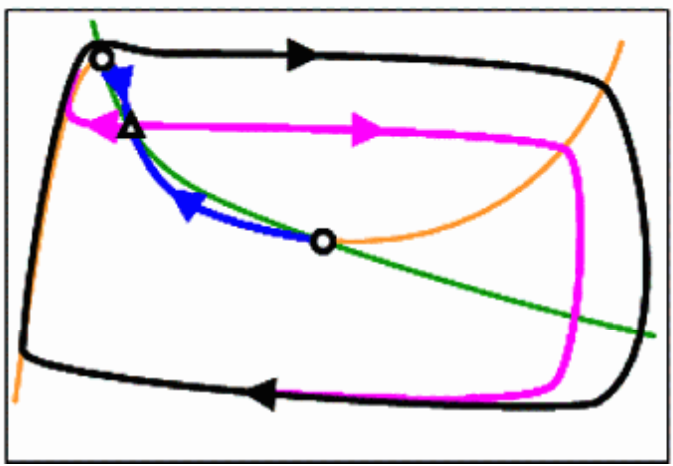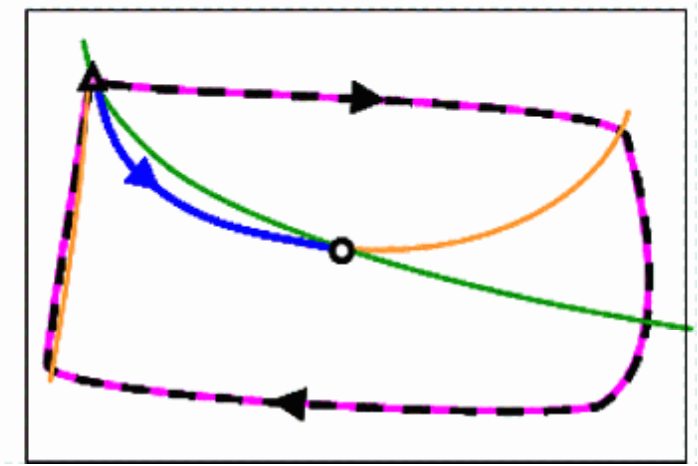

Figure 8

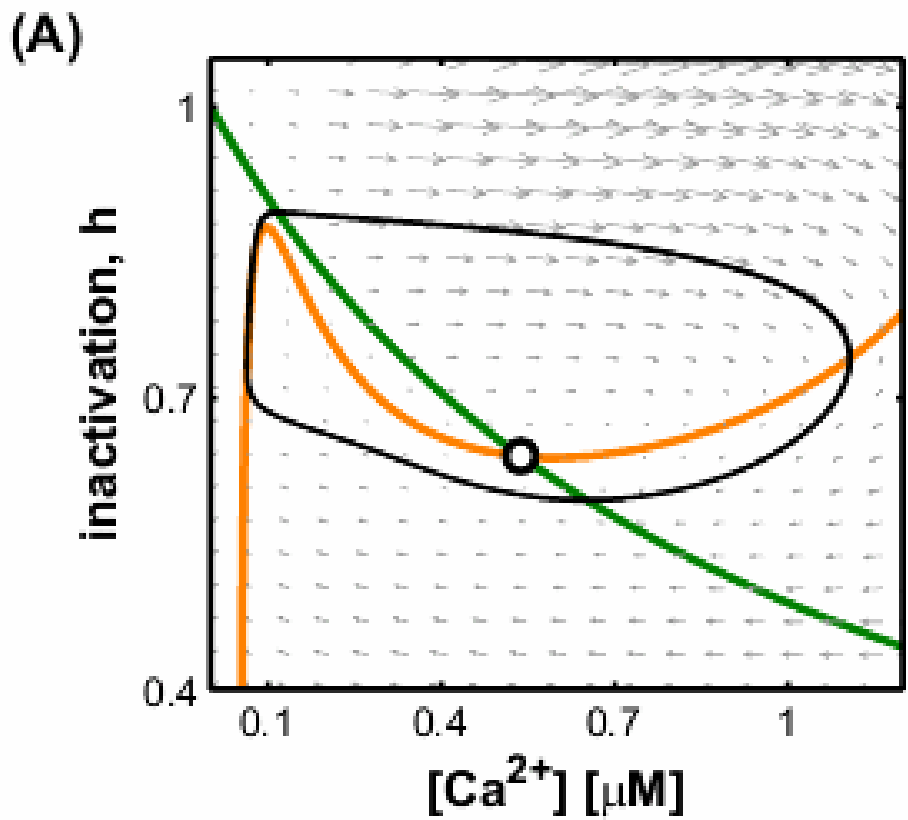 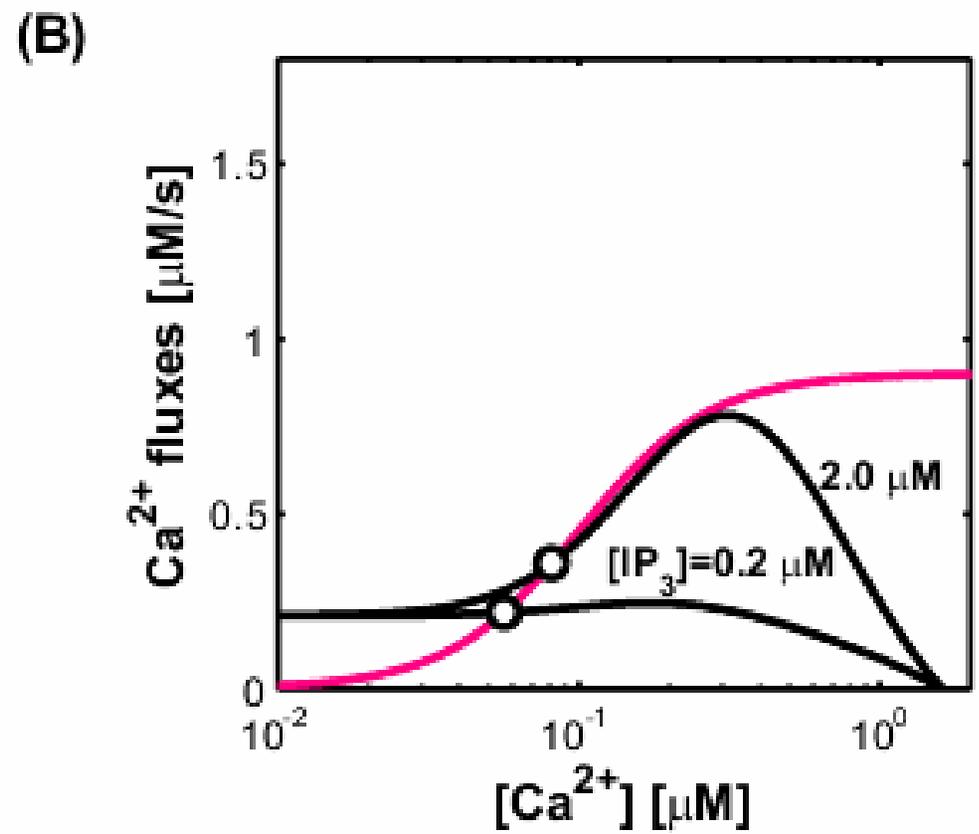

Figure 9

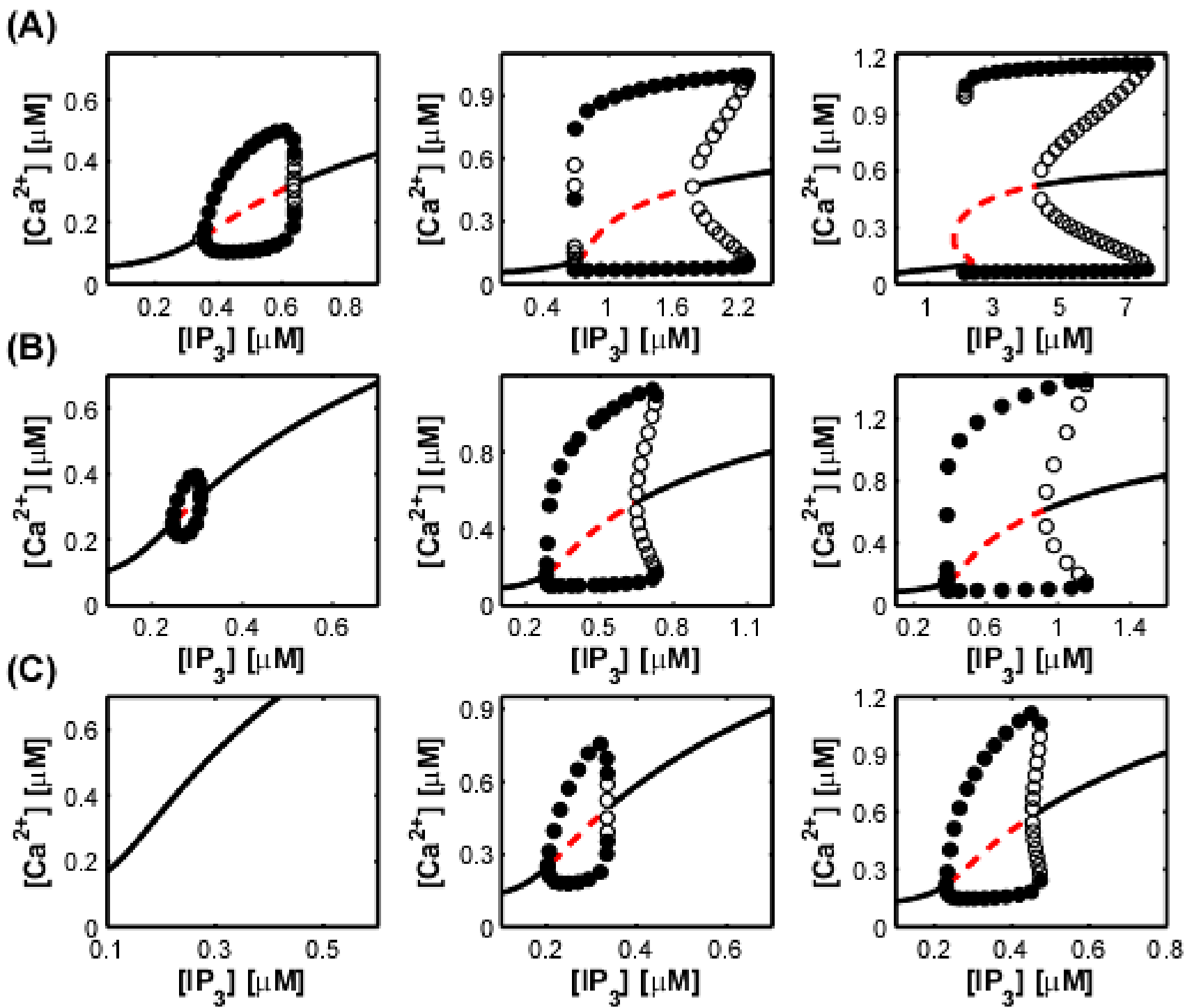

Figure 10

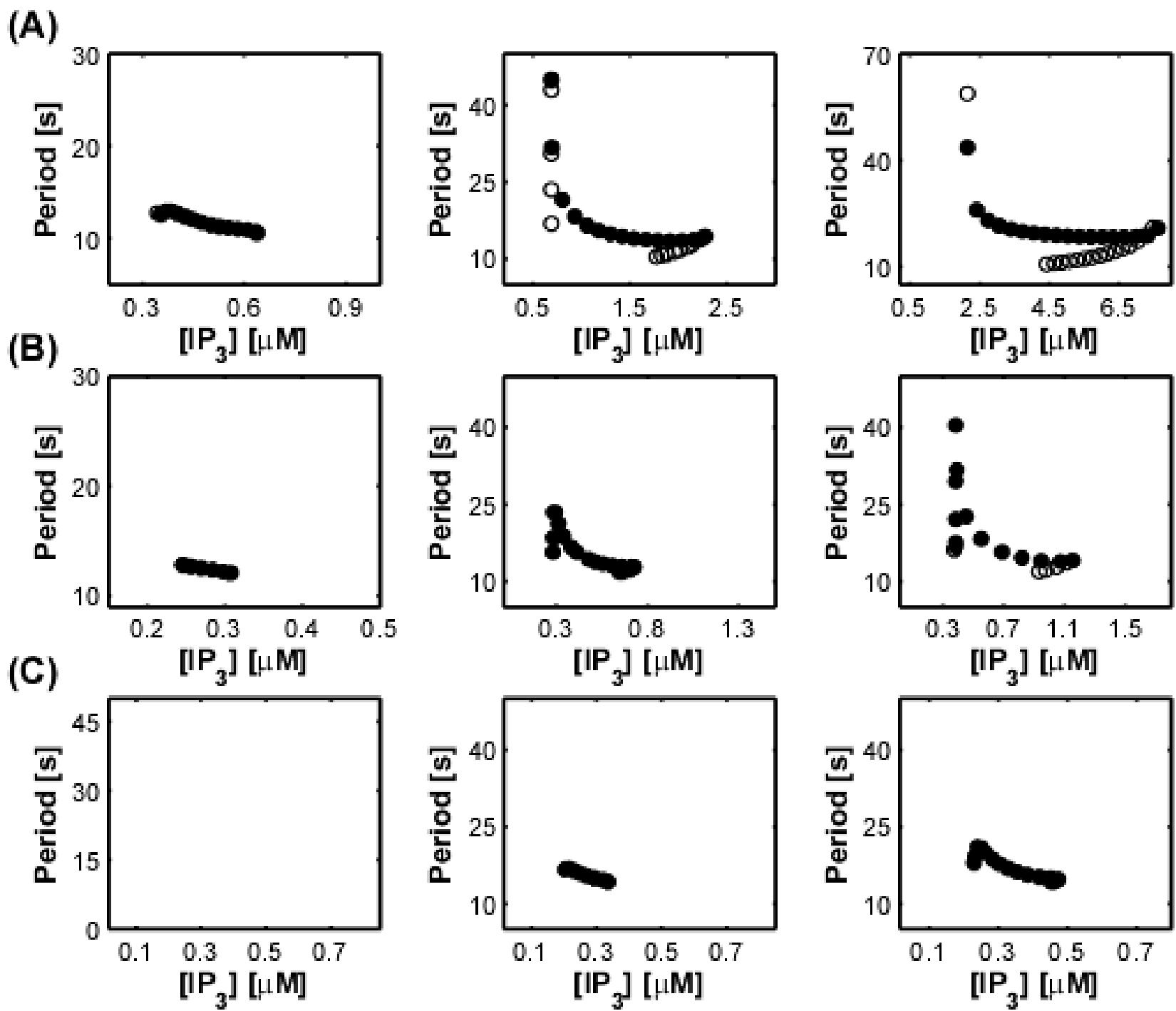

Figure 11

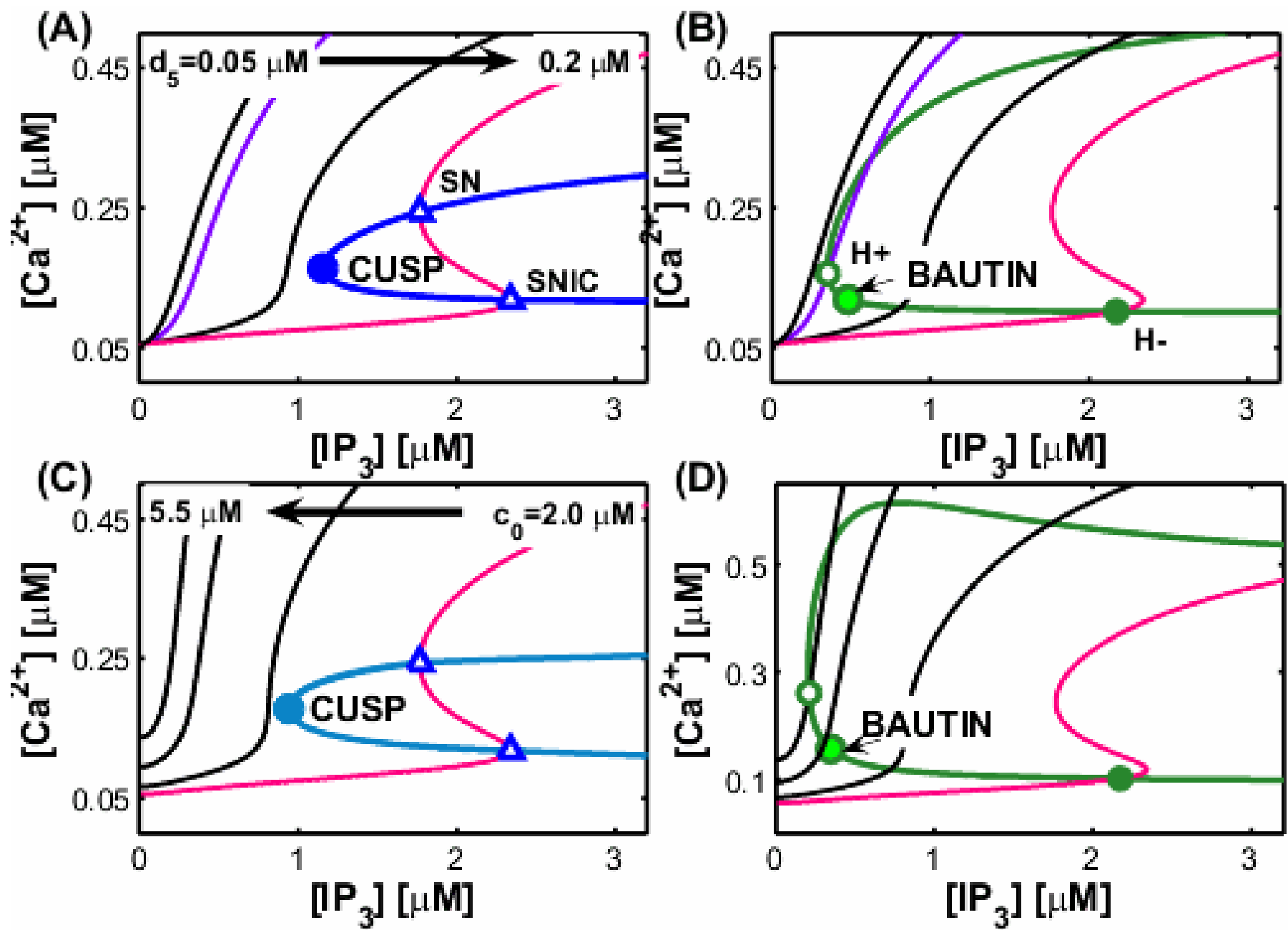

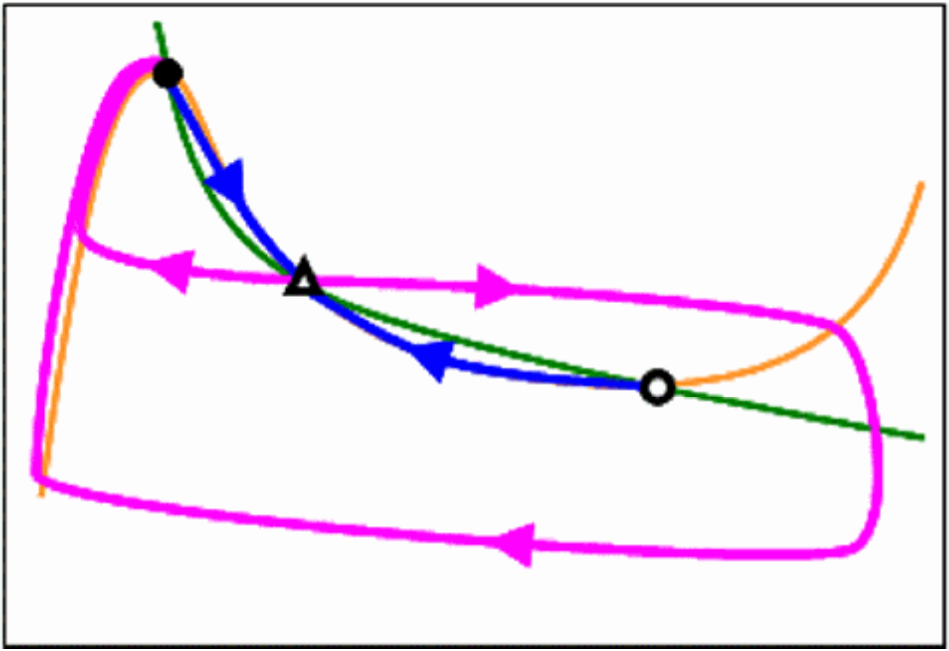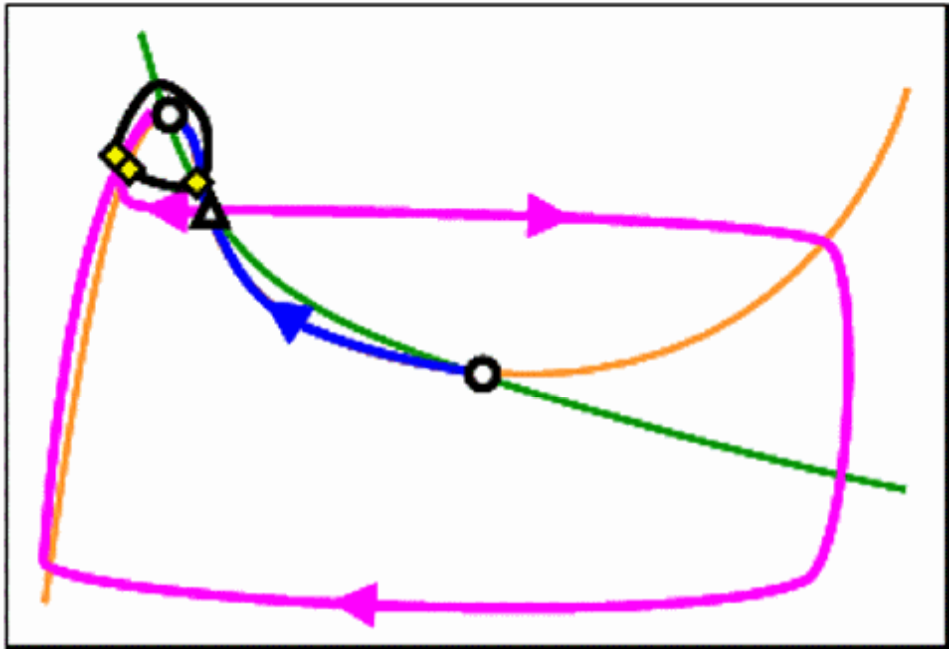

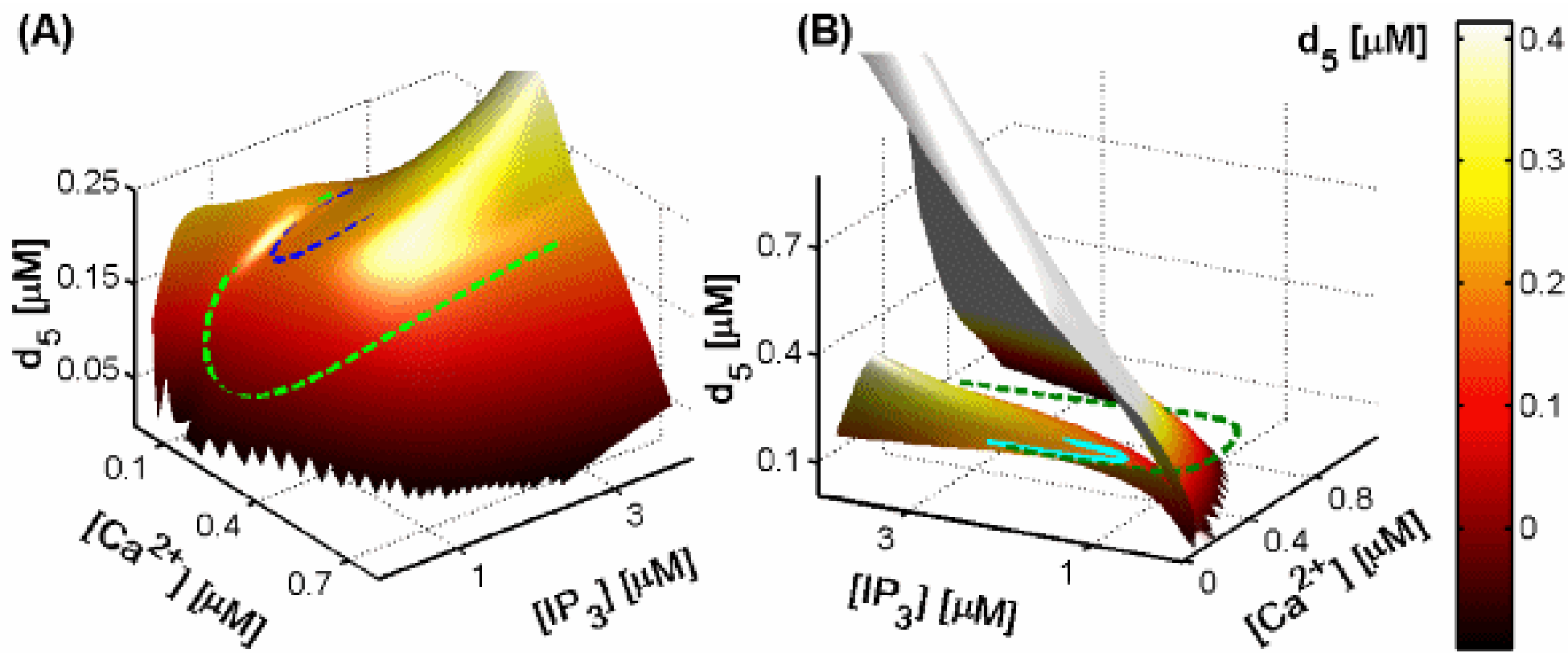

Figure 14

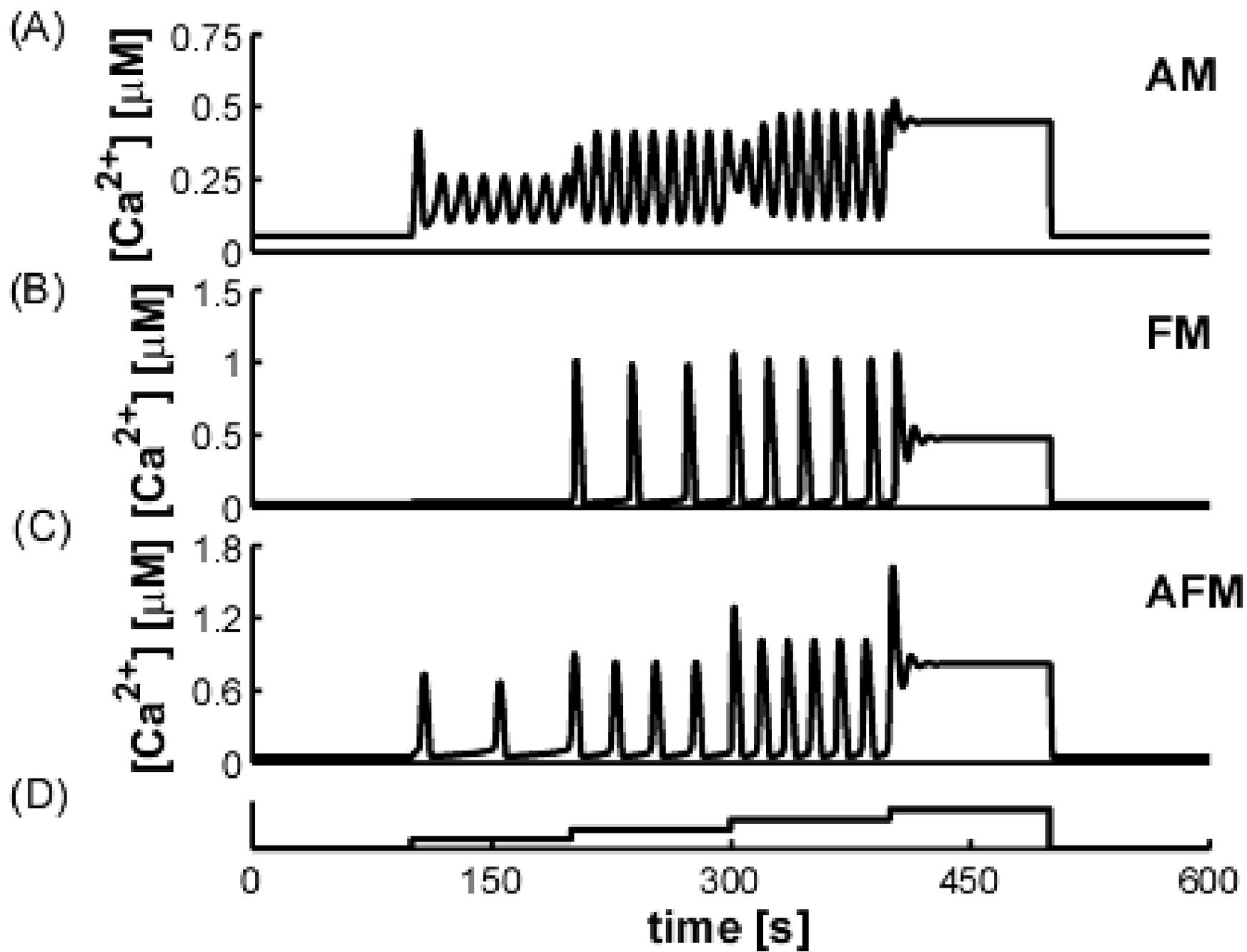

Figure 15

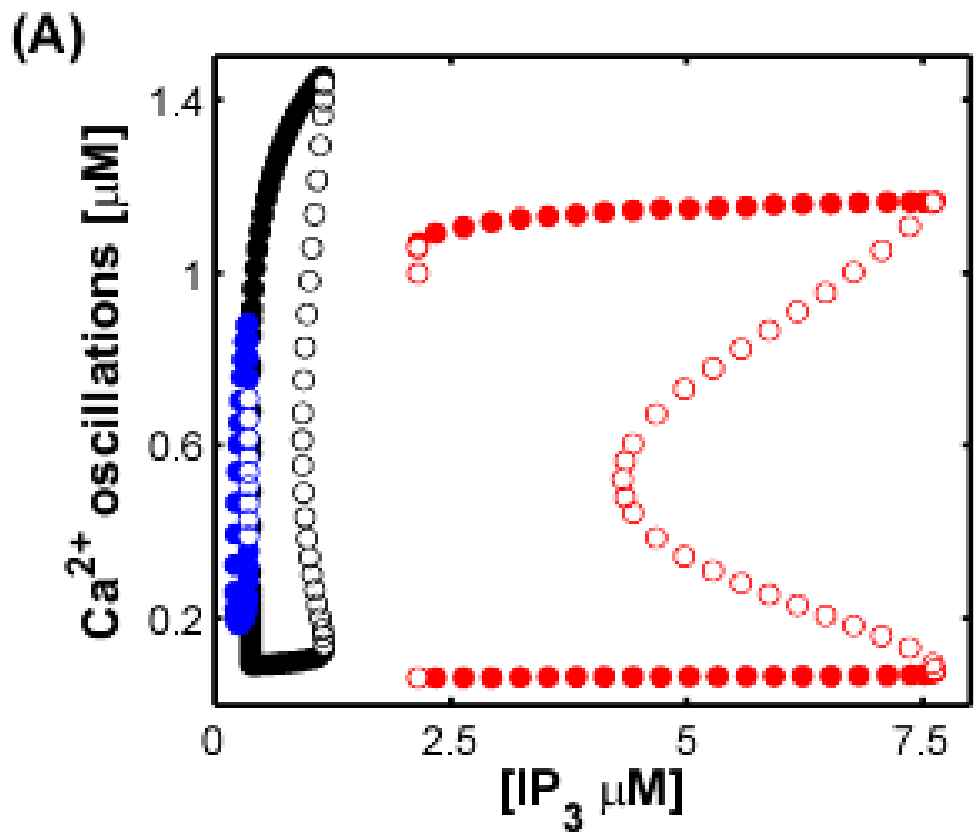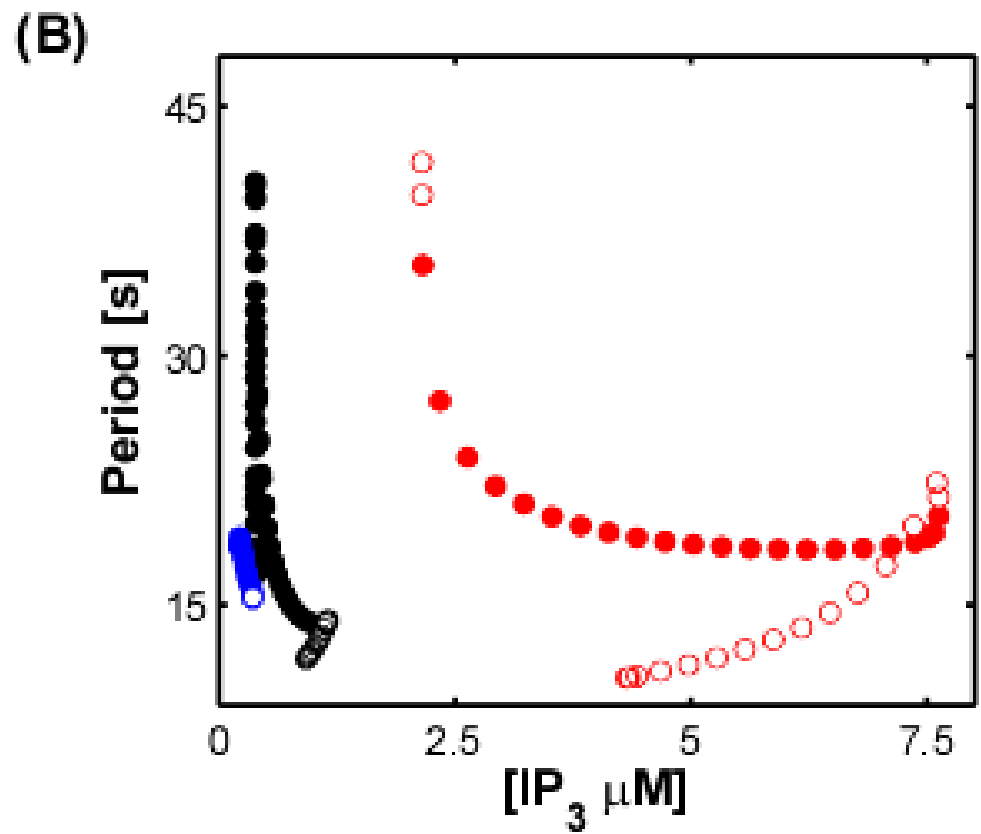

### (A) AM

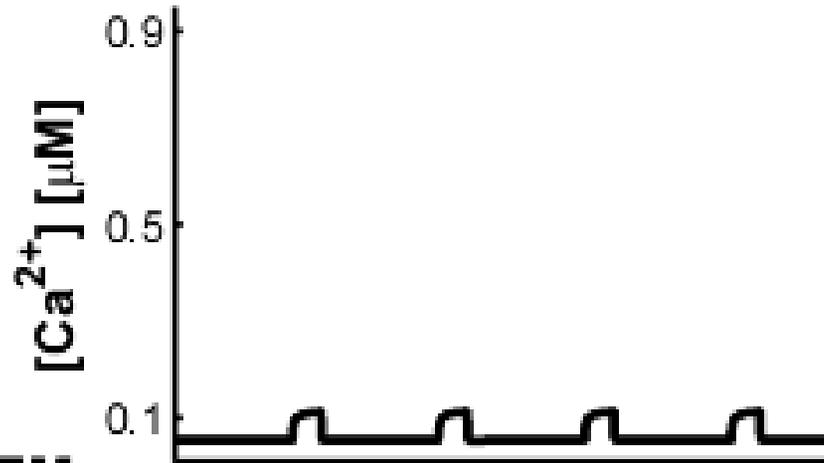
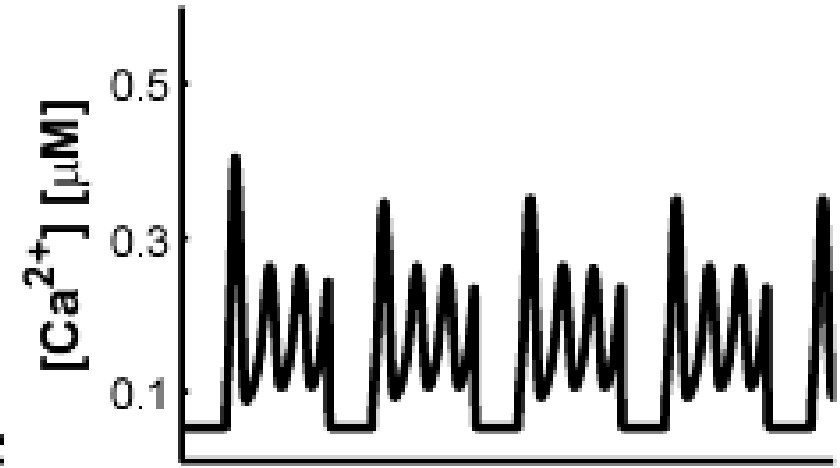

### (B) FM

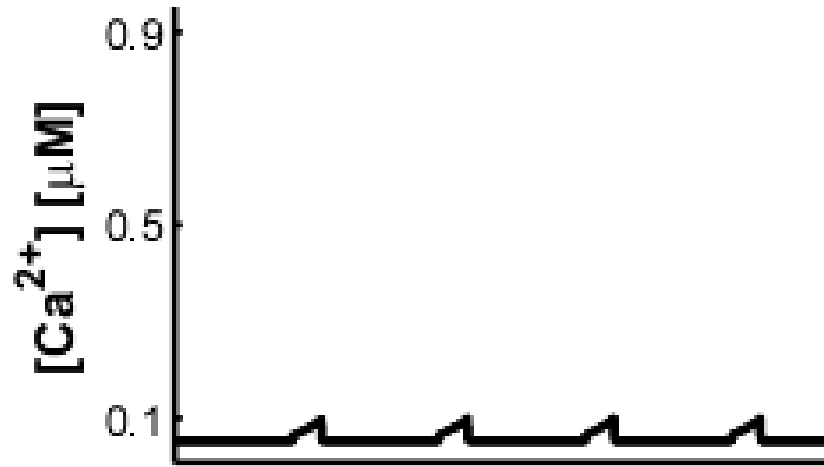
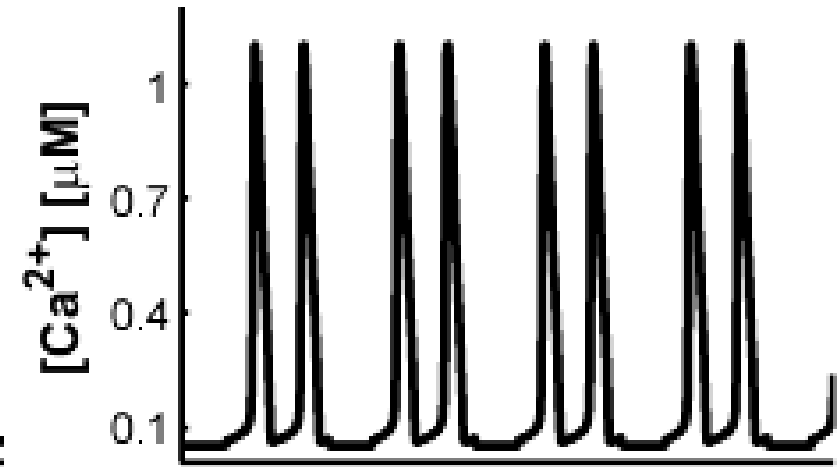

### (C) IP$_3$ STIMULUS

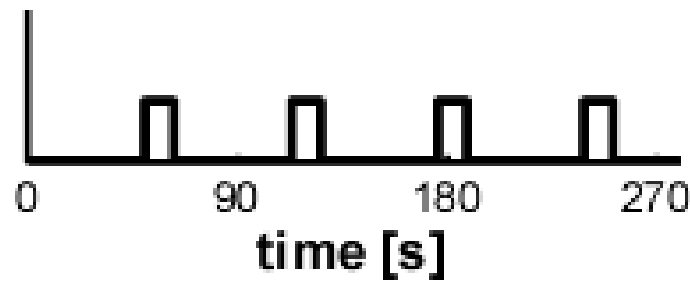
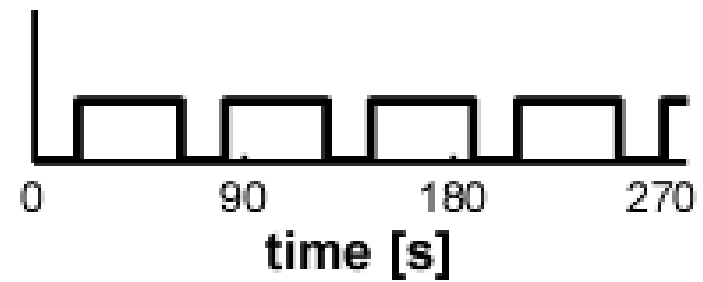